\documentclass{ametsocV6.1}

\title{Climate impacts of forced equatorial superrotation in an idealized GCM}

\authors{Tim Marino\aff{a}\correspondingauthor{Tim Marino, tim.marino@ens-lyon.fr}, Michael P. Byrne\aff{b}, Corentin Herbert\aff{c}}

\affiliation{\aff{a}{ENS de Lyon, CNRS, LPENSL, UMR 5672, 69342, Lyon cedex 07, France}\\
\aff{b}{School of Earth and Environmental Sciences, University of St Andrews, St Andrews, UK}\\
\aff{c}{CNRS, ENS de Lyon, LPENSL, UMR 5672, 69342, Lyon cedex 07, France}}

\abstract{While it is expected that the large-scale tropical circulation should undergo some changes in a warmer climate, it remains an open question whether its characteristic features, such as the Hadley cell, the intertropical convergence zone, or the weak surface easterlies, could take a completely different shape.
  As an example, it has been hypothesized that the Earth’s atmosphere may have experienced equatorial superrotation – i.e. westerly winds at the equator – during its history.
  The possibility of equatorial superrotation has been studied in a range of planetary atmospheres, including Earth-like ones, with the objective of understanding the underlying dynamical processes.
  However, the broader impact that this dramatic circulation change would have on the climate system is practically unexplored.
  This is the question we address here.
  We perform idealized GCM simulations with an imposed equatorial torque to investigate how a forced superrotating atmosphere affects surface temperature and the water cycle.
  We show that these effects are quite large and directly related to the global circulation changes, which extend beyond the tropical atmosphere.
  Using tools including a forcing/feedback analysis and a moist energy balance model, we argue that the dominant mechanism is changes in atmospheric energy transport, driven in particular by the collapse of the meridional overturning circulation, and to a smaller extent by the appearance of an equatorial jet, and the concomitant redistribution of moisture in the tropics, leading to a much weaker relative humidity gradient which has strong radiative effects.}

\begin{document}

\maketitle

\section{Introduction}

In the context of anthropogenic climate change, the response of the atmospheric circulation to radiative forcing is often seen as a second-order effect, even if it is widely acknowledged that it can have large consequences localized in time and space, for instance for regional climate and extreme events~\citep[e.g.]{Cohen2014, Shepherd2014, Vautard2023}.
In principle, it is possible that in a warmer climate a major reorganization of the large-scale circulation would drive further global changes, either through positive feedbacks acting uniformly in space or by changing the energy budget of large portions of the globe.

While fundamental balance relations enforce strong constraints on the large-scale mid-latitude circulation, dramatic changes are not \emph{a priori} excluded for the tropical circulation, and it has been suggested that major surprises might occur there, potentially through coupling with the ocean~\citep{pierrehumbertClimateChangeTropical2000}.
While current projections do exhibit changes for instance in the extent of the Hadley cell~\citep{Seidel2008}, the intertropical convergence zone~\citep{Byrne2018}, and a general weakening of the overturning circulation~\citep{Merlis2015, Chemke2023}, they do not foresee major changes in the structure of the circulation. On the other hand, the emergence of equatorial superrotation -- the reversal of Earth's climatological easterly winds in the tropics~\citep{heldEquatorialSuperrotationEarthlike1999} -- would constitute a fundamental change in the atmospheric general circulation. Superrotation has long been studied by atmospheric dynamicists and modelers~\citep{suarezTerrestrialSuperrotationBifurcation1992,saravananEquatorialSuperrotationMaintenance1993,heldEquatorialSuperrotationEarthlike1999}, but has attracted renewed interest in recent years partly because of its relevance for planetary atmospheres~\citep{readSuperrotationVenusTitan2018, Imamura2020}.
The suggestion that equatorial superrotation might have occurred during warm climates of Earth's past, such as those of the Pliocene~\citep{tzipermanPlioceneEquatorialTemperature2009b}, Early Eocene~\citep{caballeroSpontaneousTransitionSuperrotation2010}, or even deeper in time in the Mesozoic~\citep{lanWeakEquatorialSuperrotation2023}, has further motivated study of this atmospheric phenomenon.

Until now, the focus of superrotation research has been on the dynamical processes sustaining equatorial westerlies, as this requires non-trivial anti-diffusive angular momentum transport mechanisms~\citep{heldEquatorialSuperrotationEarthlike1999}.
Several routes have been suggested, all involving momentum transport by equatorial waves, either through enhanced diabatic heating~\citep{Kraucunas2005, arnoldAbruptTransitionStrong2012, Lutsko2018} or through a hydrodynamic instability~\citep{Williams2003, Mitchell2010, Potter2014, Zurita-Gotor2018a, Zurita-Gotor2022, Zurita-Gotor2024}.
It has further been suggested that those mechanisms might include positive feedbacks leading to an abrupt transition~\citep{arnoldAbruptTransitionStrong2012, herbertAtmosphericBistabilityAbrupt2020, Zurita-Gotor2022}, which might make superrotation an example of a tipping point.
Most of these dynamical studies were conducted in an idealized modeling framework of a dry atmosphere without explicitly taking into account processes including condensation of water vapor and radiation. Two exceptions are the studies of~\citet{caballeroSpontaneousTransitionSuperrotation2010}, which observed a spontaneous transition to superrotation under large radiative forcing, and~\citet{arnoldEnhancedMJOlikeVariability2013}, which reported weak superrotation in response to increasing sea surface temperature (SST).
\citet{CaballeroCarlson2018} also investigated the plausibility of the scenarios suggested by~\citet{pierrehumbertClimateChangeTropical2000} and~\citet{tzipermanPlioceneEquatorialTemperature2009b} by studying the possibility that tropical westerlies reached the surface.

In this work, rather than focusing on the dynamical processes underlying superrotation, we instead explore the consequences that such a large-scale circulation change would have for climate if it were to occur, in particular for surface temperature and the water cycle.
To do so, we impose superrotation using a prescribed torque in a moist model which includes parameterizations for major physical processes in the atmosphere such as radiation, convection, and condensation (see Section~\ref{sec:model} for details).
We find that superrotation is not restricted to a zonal wind change in the upper tropical troposphere but rather affects the general circulation as a whole, resulting in substantial impacts on surface temperature and precipitation.
These impacts could potentially be important for interpreting paleoclimate proxies and the processes controlling warm climates in Earth's past and future. The impact of (forced) superrotation on surface temperature, revealed in this study, together with the emergence of superrotation in simulations of hot climates ~\citep{caballeroSpontaneousTransitionSuperrotation2010,arnoldEnhancedMJOlikeVariability2013} suggests a possible feedback between superrotation and SST that has received little attention to date.

After presenting the modeling setup used in our study (Section ~\ref{sec:model}), we discuss the temperature and precipitation changes induced by superrotation (Section ~\ref{sec:changes}).
We then build an understanding of the physical processes underlying the temperature changes by investigating the top-of-atmosphere (TOA) radiative budget using a forcing/feedback framework (Section ~\ref{sec:forcing/feedback}) and reconstructing the simulated temperature response using an energy balance model (Section ~\ref{sec:ebm}). Finally, we relate the superrotation-induced water cycle changes to the circulation changes (Section ~\ref{sec:p-e}) before summarizing our results (Section ~\ref{sec:conc}).

\section{Methodology} \label{sec:model}

\subsection{Numerical experiments}

In this study, we use the GCM \emph{Isca}~\citep{vallisIscaV10Framework2018} in a relatively idealized moist aquaplanet configuration.
The model, based on the GFDL dynamical core~\citep{Gordon1982}, solves the moist primitive equations in vorticity-divergence form on 40 (unevenly spaced) $\sigma$ levels~\citep{Bourke1974} at T42 horizontal resolution.

Latent heat of condensation and the radiative effects of water vapor are simulated, with the former represented through the large-scale condensation parameterization of~\cite{Frierson2006}. A standard Betts-Miller scheme for moist convection is also employed~\citep{bettsNewConvectiveAdjustment1986,bettsNewConvectiveAdjustment1986a}.
Radiative fluxes are computed using the comprehensive SOCRATES scheme~\citep{manners2017socrates}, with a CO\textsubscript{$2$} concentration of $300$~ppm and a modern ozone climatology.
There is a diurnal cycle in the model but no seasonal cycle; a permanent equinox configuration is used for insolation.
Clouds are not represented in our configuration of the model and the radiative effect of aerosols is not taken into account.

The lower boundary is a global slab ocean with a $20$~m mixed layer depth.
We impose prescribed meridional heat fluxes in the ocean, following~\cite{Merlis2013}.
Sea ice is not represented in the model and the surface temperature is allowed to reach values below the freezing point.
The surface albedo is prescribed to $0.25$ and standard drag laws govern the exchanges of momentum, sensible and latent heat at the surface, based on Monin-Obukhov theory as described in~\cite{Frierson2006}. 
At the top of the atmosphere (above $150~\rm{hPa}$), a sponge layer dissipates momentum with a Rayleigh drag of characteristic timescale $12~\rm{h}$. 

This type of idealized configuration (with variations regarding, for example, how radiative transfer is treated or whether prescribed or interactive SSTs are used) has been used extensively, not only as an evaluation framework for model intercomparison~\citep{Neale2000, Neale2000a, Lee2008} but also as a conceptual framework to connect theory with numerical models which remain interpretable~\citep{Blackburn2013}.
For instance, similar configurations have been used to study meridional energy transport~\citep{Caballero2005,Frierson2007c,Rose2012, Merlis2022} and the water cycle~\citep{OGorman2008,Byrne2016,Donohoe2019} over a range of climates, the response of the Hadley cells to various forcings~\citep{Levine2011, Merlis2013, hilgenbrinkResponseHadleyCirculation2018}, convectively coupled waves~\citep{Frierson2007}, and the annual cycle of the tropical tropopause layer~\citep{Jucker2017}.

Here, we use the idealized configuration described above because it constitutes a relatively simple framework that nevertheless takes into account key feedbacks linking surface temperature and the water cycle to changes in large-scale circulation.
The strategy we adopt to investigate the response of climate-relevant fields to the emergence of superrotation is to impose an atmospheric torque.
We choose this numerical setup, rather than, for instance, letting superrotation emerge spontaneously from thermal forcings such as increased CO\textsubscript{2} or sea surface temperatures, for two reasons.
On the one hand, although the spontaneous emergence of superrotation has been reported for such thermal forcings in a small number of studies~\citep{caballeroSpontaneousTransitionSuperrotation2010, arnoldEnhancedMJOlikeVariability2013,lanWeakEquatorialSuperrotation2023}, the transition occurs for very large forcing values and might be sensitive to parameterization choices~\citep{Zurita-Gotor2025}.
Hence it remains unclear to what extent it is robust and well represented by current models.
In addition, our goal here is to isolate the climate changes, in particular the temperature and precipitation responses, resulting directly from the circulation changes associated with the transition to superrotation.
This is more easily achieved with a forcing which does not directly affect these fields.

With this in mind, we impose a torque with a spatial structure corresponding to the convergence of the eddy zonal momentum fluxes generated by a zonally asymmetric diabatic heating, which has been used in a number of dynamical studies of superrotation~\citep{Kraucunas2005, arnoldAbruptTransitionStrong2012, Lutsko2018, herbertAtmosphericBistabilityAbrupt2020}.
This spatial structure of this imposed torque is the following:
\begin{align}
    \mathbf{F} = F_0 \frac{\left( 1 - \left(\frac{\phi}{\phi_{\mathrm{scale}}}\right)^2\right)}{\cos{\phi}} \exp\left(-\frac{1}{2}\left( \frac{\phi}{\phi_{\rm scale}} \right)^2 \right) \exp\left( -\frac{1}{2}\left( \frac{p-p_0}{p_{\rm scale}}\right)^2\right) \mathbf{e}_x,
\end{align}
where $\phi_{\rm scale} = 5^{\circ}$, $p_0=300$~hPa, $p_{\rm scale} = 100$~hPa and $\mathbf{e}_x$ is the zonal unit vector.
The torque has a tripolar structure, which is also similar to observed eddy momentum flux convergence associated to seasonal superrotation on Earth~\citep{Zhang2022}.
This structure integrates to $ \pi \exp\left(-\frac{1}{2}\frac{\pi^2}{4\phi_{\rm scale}^2}\right) \approx 0$ on the meridional direction, which guarantees no net injection or damping of angular momentum. 
We vary the strength of the prescribed torque through the $F_0$ parameter, from $0$ to $4$~m.s\textsuperscript{$-1$}.day\textsuperscript{$-1$}.
This parameter corresponds to the maximum value of the torque.
While we use it to identify the different simulations, it should be kept in mind that typical values over regions of finite extent are smaller and similar to accelerations associated with the transition to superrotation reported elsewhere in the litterature.
Note that eddy momentum flux convergence leading to superrotation could also be generated, for example, by a low-latitude mixed barotropic-baroclinic instability~\citep{Williams2003}, or by a Kelvin-Rossby instability~\citep{Mitchell2010, Potter2014, Zurita-Gotor2018a, Zurita-Gotor2022, Zurita-Gotor2024}, leading to eddy fluxes with a similar shape, albeit less focused on the tropics.

\subsection{Forcing/feedback framework} \label{subsec:forcing/feedback}

We run several experiments with the model described above and interactive SSTs: a control run, with $F_0=0$~m.s\textsuperscript{$-1$}.day\textsuperscript{$-1$}, as well as forced runs with unit increments of $F_0$ up to $F_0=4$~m.s\textsuperscript{$-1$}.day\textsuperscript{$-1$} (we will mostly focus on the $F_0 = 0$ and $F_0 = 4$~m.s\textsuperscript{$-1$}.day\textsuperscript{$-1$} experiments).
Analogous runs but with fixed SSTs are also performed: in those cases, we prescribe time-averaged SSTs from the control run.
The fixed-SST experiments are run in order to diagnose the radiative forcing induced by the prescribed superrotation, in the standard way described for instance by~\citet{Dessler2015}.
In particular, our aim is to decompose the changes in TOA radiation into radiative forcing and feedback components, which can each be broken down into contributions from different physical mechanisms.
In this context, we can write the anomalous TOA energy budget as:
\begin{align} \label{eq:forcing/feedback}
    \underbrace{\delta R}_{\text{Total TOA budget}} = \underbrace{F}_{\text{Radiative forcing}} - \underbrace{\lambda \delta T_{\rm s}}_{\text{Feedbacks}},
\end{align}
where $\lambda$ is the climate feedback parameter.

The radiative forcing term $F$ in Eq.~(\ref{eq:forcing/feedback}), diagnosed using the fixed-SST simulations, can be interpreted as the change in the TOA radiative budget driven by the prescribed torque and resulting circulation change.
The feedback term $\lambda \delta T_{\rm s}$, on the other hand, includes all changes in TOA radiation linearly related to surface temperature.
This is similar to the standard framework used for studying climate sensitivity~\citep{sherwoodAssessmentEarthsClimate2020}, with one notable difference.
In the context of climate-sensitivity studies, the radiative forcing is driven only by the direct radiative effect of CO\textsubscript{2} changes, with temperature and humidity held constant (in practice, estimating this radiative forcing precisely can be challenging due to `rapid adjustments' of, for example, humidity to CO\textsubscript{2} changes that are difficult to disentangle from the direct CO\textsubscript{2} forcing \citep[e.g.,][]{smith2018understanding}).
In this study, the forcing does not directly affect the TOA radiative budget; instead the temperature and humidity profiles first adjust to the circulation changes at essentially constant SST, leading to a modified TOA radiative budget which we term the radiative forcing (this is analogous to the \emph{effective radiative forcing} in the context of CO\textsubscript{2} forcing \citep[e.g.,][]{smith2020effective}).
The slower adjustment of SST leads to further reorganization of the water vapor and temperature fields, which are included in the radiative feedback terms.

\subsection{Radiative kernels}\label{sec:radiative_kernel}

We decompose changes in the TOA radiation budget into components representing different physical processes using radiative kernels~\citep{sodenQuantifyingClimateFeedbacks2008}.
Radiative kernels represent the change of the radiative budget associated to a unit change of a given quantity at a given point in space.
In general terms and assuming linearity, we can approximately write a TOA radiation change, $\delta R$, as a function of the radiative kernels and changes in the various physical quantities affecting TOA radiation: $\delta R \approx \sum_X K^X \delta X$, where $K^X=\partial R/\partial X$ is a radiative kernel and $X$ represents quantities affecting the TOA radiation.
In the configuration considered here, there are three relevant radiative kernels, $K^{T_{\mathrm{s}}}  = \left.\frac{\partial R}{\partial T_{\mathrm{s}}}\right|_{T,q}$, $K^{T} = \left.\frac{\partial R}{\partial T}\right|_{T_{\mathrm{s}},q}$ and $K^{\mathrm{wv}} = \left.\frac{\partial R}{\partial \log(q)}\right|_{T_{\mathrm{s}},T} \frac{\partial \log(q)}{\partial T}$, quantifying the sensitivity of the TOA radiative budget to changes in surface temperature, atmospheric temperature and specific humidity, respectively (there are no changes in chemical species concentrations or albedo in our simulations).

In the analysis below, we will use these kernels to decompose the changes of TOA radiative budget, but we will use relative humidity as a variable instead of specific humidity, so that the total TOA radiative budget changes write $\delta R_{\mathrm{tot, kernel}} = \delta R_{T_{\mathrm{s}}}+\delta R_{\Gamma}+\delta R_{\mathrm{RH}}$, with $\delta R_{T_{\mathrm{s}}}=( K^{T_{\mathrm{s}}} + K^{T} + K^{\mathrm{wv}}) \delta T_{\mathrm{s}}$, $\delta R_{\Gamma}=( K^T + K^{\mathrm{wv}}) (\delta T - \delta T_{\mathrm{s}})$ and $\delta R_{\mathrm{RH}}=K^{\mathrm{wv}} \frac{\delta \log\left(\mathrm{RH}\right)}{\partial \log(q) / \partial T}$ (see Appendix~B for a derivation).

In practice, radiative kernels are computed by running a radiation scheme on two base states that differ only  by a local change in the chosen variable, for instance a $1$~K temperature change or a $1$~\% specific humidity change at a given point in space.
Here, we use kernels previously computed for Isca using the same setup as that employed in this study (aquaplanet, no clouds or sea ice, T$42$ resolution, $40$ vertical levels)~\citep{liuRadiativeKernelsIsca2020}, but with the RRTM radiative scheme~\citep{mlawerRadiativeTransferInhomogeneous1997} (rather than SOCRATES).
To validate this approach, we have verified that the total radiative budget computed from the kernels approximately matches the simulated radiative budget from the experiments (see Section~\ref{sec:forcing/feedback}.\ref{sec:radiative_kernel_rh}).

Note that, as mentioned above, unlike the more familiar case of CO\textsubscript{2} forcing, here both the radiative forcing and the radiative feedback terms introduced in Section~\ref{sec:model}\ref{subsec:forcing/feedback} can be decomposed using the radiative kernels: we can write $\delta R = \sum_X \delta R_X = F-\lambda \delta T_s=\sum_X F_X-(\lambda \delta T_S)_X$, where the sum runs over the surface temperature, air temperature and humidity fields.

\section{Climate responses to superrotation}\label{sec:changes}

In this section, we describe the impact of the imposed torque on the atmospheric circulation (Section~\ref{sec:changes}\ref{sec:circulation_changes}) and on climate, in particular surface temperature (Section~\ref{sec:changes}\ref{sec:tsurf_changes}) and the distribution of precipitation minus evaporation (Section~\ref{sec:changes}\ref{sec:precip_changes}). The physical mechanisms associated with these simulated changes will be discussed in more detail in Sections~\ref{sec:forcing/feedback},~\ref{sec:ebm} and~\ref{sec:p-e}.

\subsection{Atmospheric circulation}\label{sec:circulation_changes}

\begin{figure}
    \centering
  \includegraphics[width=36pc]{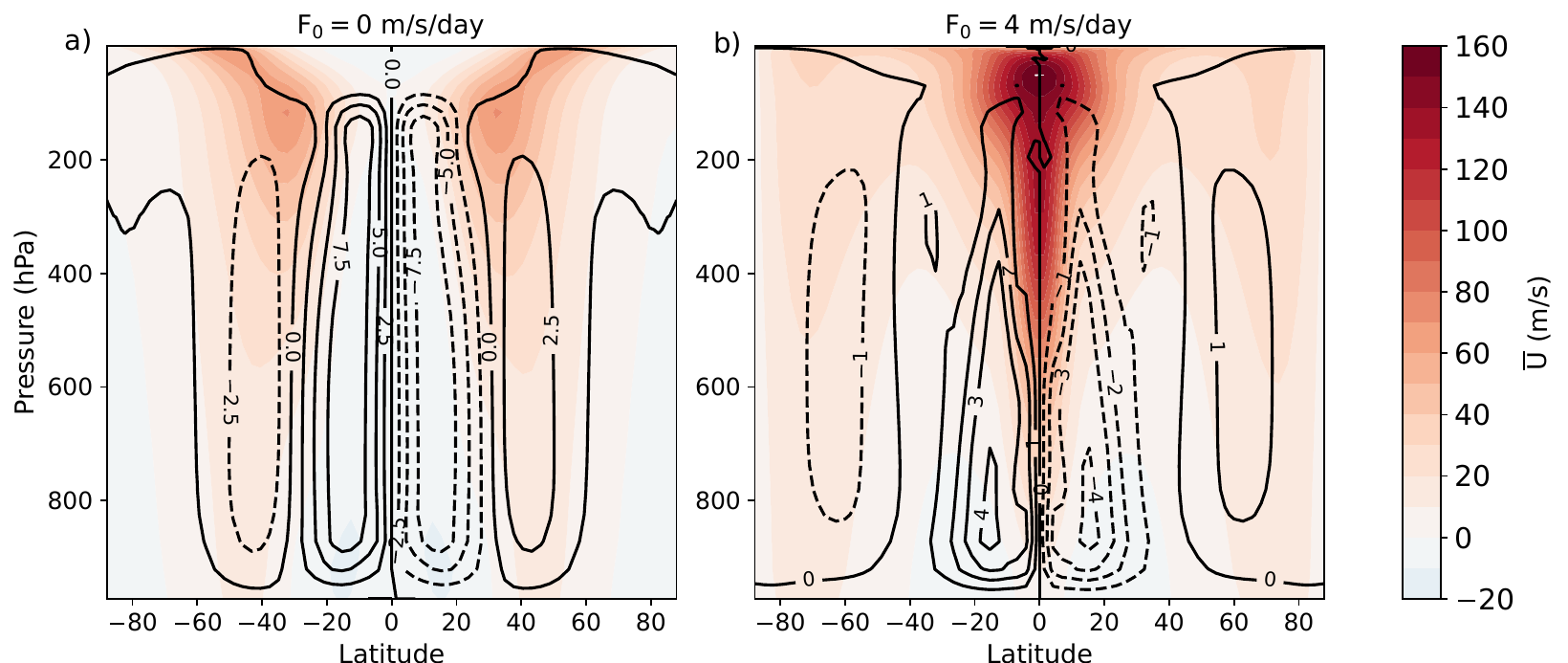}\\
  \caption{Zonally- and time-averaged zonal wind (colors) and meridional mass streamfunction (contours with an interval of $10^{10}~\mathrm{kg.s}^{-1}$ and dashed lines indicating negative values), for experiments with $F_0 = 0$~m.s\textsuperscript{$-1$}.day\textsuperscript{$-1$} (control run, panel \textbf{a}) and $F_0 = 4$~m.s\textsuperscript{$-1$}.day\textsuperscript{$-1$} (panel \textbf{b}) and interactive SSTs. }\label{fig:jet}
\end{figure}

We first consider the response of the time-averaged atmospheric circulation to the imposed upper-tropospheric torque.
Figure~\ref{fig:jet} shows the zonally-averaged zonal wind and the meridional mass streamfunction for both the control run ($F_0=0$~m.s\textsuperscript{$-1$}.day\textsuperscript{$-1$}, panel \textbf{a}) and the run with the largest prescribed torque ($F_0=4$~m.s\textsuperscript{$-1$}.day\textsuperscript{$-1$}, panel \textbf{b}).

The control run (Fig.~\ref{fig:jet}, panel \textbf{a}) exhibits the main characteristics of Earth's current atmospheric circulation, with some discrepancy due to the idealized setup.
The meridional circulation is reasonably close to annual-mean observations, except that it is hemispherically symmetric by construction.
Additionally, the two zonal jets are slightly narrower than observed or simulated in more realistic models, and the maximum wind speed in each jet is significantly stronger and reached at slightly higher altitude (this is a common property of aquaplanet experiments).

As the prescribed torque increases, we observe a transition to a state of superrotation (Fig.~\ref{fig:jet}, panel \textbf{b}). This state is characterized by westerly winds over almost the entire atmosphere (except weak easterlies in the subtropics close to the surface), and a strong zonal jet of approximately $150$~m.s\textsuperscript{$-1$} at the equator, about twice the speed of the mid-latitude jets in the control run.
The mean meridional circulation also undergoes major changes in its structure.
The Hadley cell is strongly weakened; the mass meridional streamfunction decreases by a factor $3$ in the tropics.
A residual Ferrell cell remains, albeit significantly weaker than in the control run and shifted poleward by about 20$^\circ$.
Simulations for intermediate values of $F_0$, between $1$ and $3$~m.s\textsuperscript{$-1$}.day\textsuperscript{$-1$} (not shown), demonstrate that this transition is abrupt: the strength of the equatorial jet does not increase smoothly with the forcing amplitude.
Instead, it appears suddenly at $F_0=3$~m.s\textsuperscript{$-1$}.day\textsuperscript{$-1$}.

\begin{figure}
  \centering \includegraphics[width=36pc]{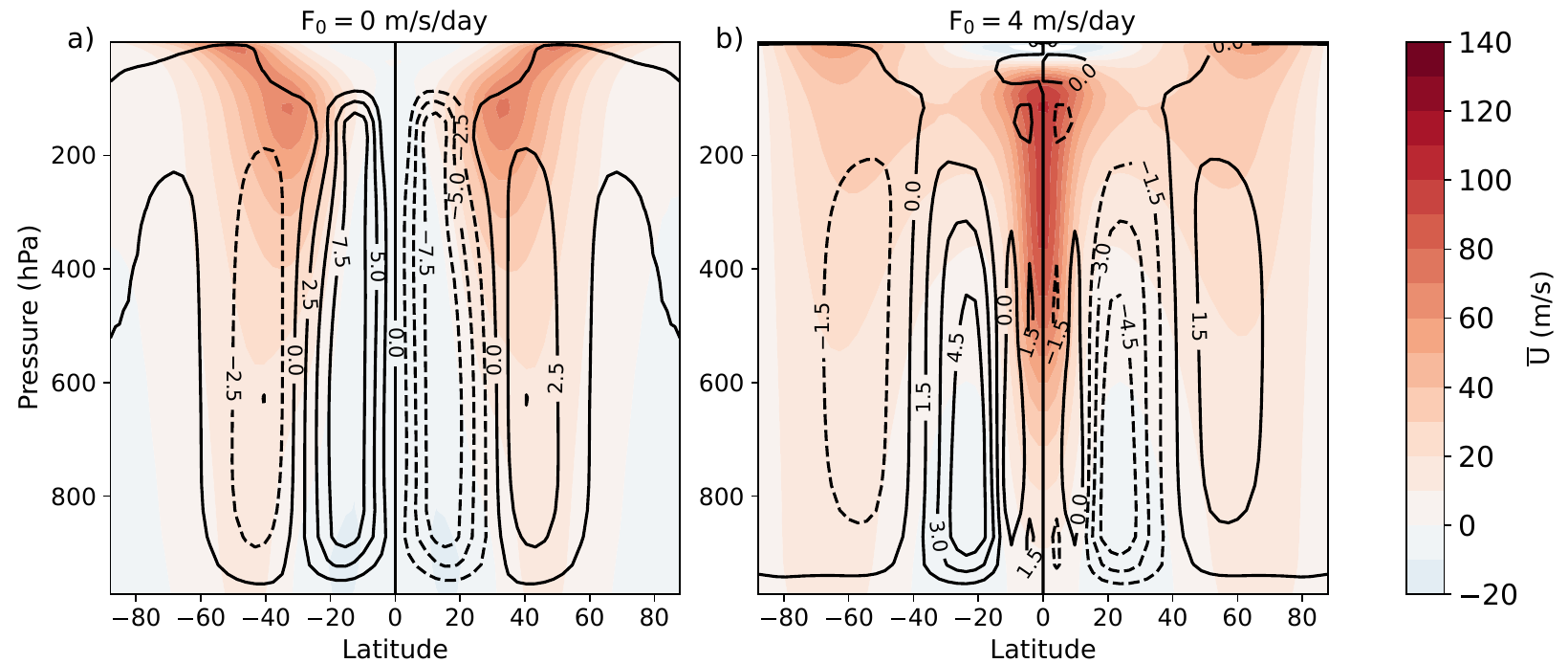}
\caption{Zonally- and time-averaged zonal wind (colors) and meridional mass streamfunction (black contours, in units of $10^{10}~\mathrm{kg.s}^{-1}$), for $F_0 = 0$~m.s\textsuperscript{$-1$}.day\textsuperscript{$-1$} (control run, panel \textbf{a}) and $F_0 = 4$~m.s\textsuperscript{$-1$}.day\textsuperscript{$-1$} (panel \textbf{b}), for the fixed SST experiments.}\label{fig:jet_fs}
\end{figure}

Similarly, the zonally-averaged atmospheric circulations for the fixed-SST experiments are shown in Figure~\ref{fig:jet_fs}.
The circulation in the control run (Fig.~\ref{fig:jet_fs}, panel \textbf{a}) is quite similar to that in the interactive-SST experiment.
As expected, we observe a similar transition to equatorial superrotation (Fig.~\ref{fig:jet_fs}, panel \textbf{b}).
However, the jet is weaker than in the interactive SST experiments, suggesting the existence of a positive feedback between surface temperature and the strength of the jet. This could be part of a self-sustaining mechanism for superrotation, should this positive feedback be strong enough to counterbalance damping in a setup with more typical forcings such as due to increases in CO\textsubscript{2} concentrations or SST.
The changes in the mean meridional circulation are also weaker but follow the same structure as in the interactive SST case: the intensities of both the direct and indirect cells decrease and they are shifted poleward.

\subsection{Surface temperature}\label{sec:tsurf_changes}

\begin{figure}[t]
    \centering
  \noindent\includegraphics[width=27pc,angle=0]{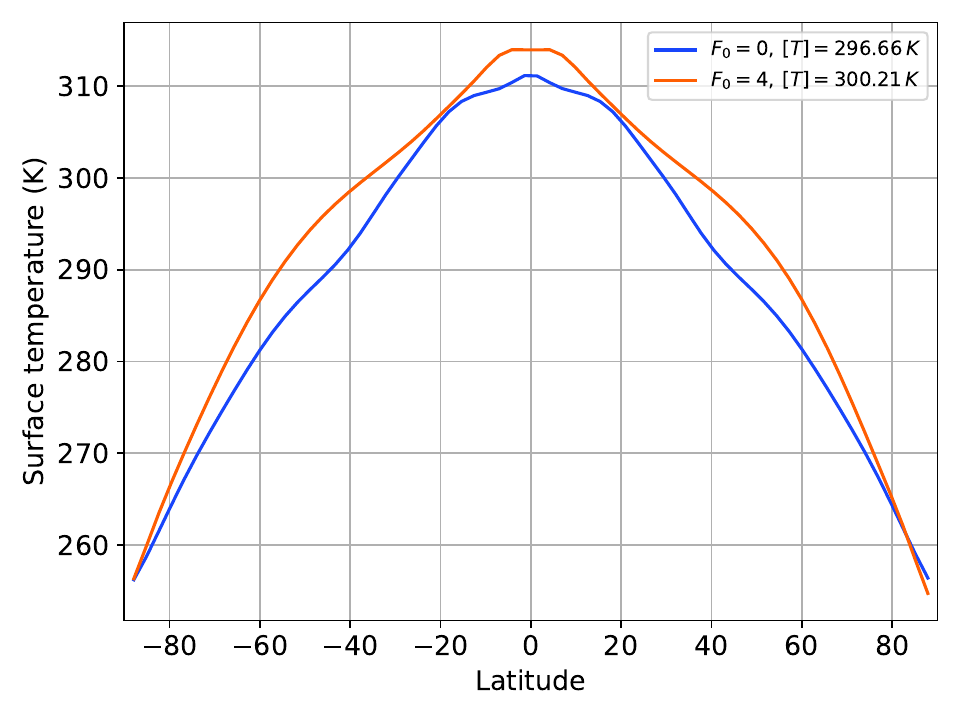}\\
  \caption{Zonally- and time-averaged surface temperature profiles for two different values of $F_0$: $F_0=0$~m.s\textsuperscript{-1}.day\textsuperscript{-1} (control run, blue curve) and $F_0=4$~m.s\textsuperscript{-1}.day\textsuperscript{-1} (superrotating state, orange curve). The value of the globally-averaged surface temperature in each case is given in the legend.}\label{fig:ts_profile}
\end{figure}

Figure~\ref{fig:ts_profile} shows the zonally- and time-averaged surface temperature profiles for the control run ($F_0 = 0$~m.s\textsuperscript{-1}.day\textsuperscript{-1}) and in a forced superrotating state ($F_0 = 4$~m.s\textsuperscript{-1}.day\textsuperscript{-1}) for the simulations with a slab ocean.
We first note that the control run is relatively warm compared to the real Earth system, particularly in the tropics, and has a strong equator-to-pole surface temperature gradient.
These features of our idealized experiments are likely related to prescribing a relatively small, spatially uniform albedo and to the lack of a seasonal cycle for insolation, which leads to warmer tropics and colder polar regions.
The globally-averaged surface temperature is nevertheless comparable to similar aquaplanet experiments~\citep[e.g.]{hilgenbrinkResponseHadleyCirculation2018}.

In the superrotating state, surface temperature increases almost everywhere.
The largest changes are in the mid-latitudes (between $30$ and $70^{\circ}$) and in the deep tropics (between $0$ and $10^{\circ}$).
In the subtropics (around $20^{\circ}$), the surface temperature changes induced by forced superrotation almost vanish.
The globally-averaged surface temperature increases by approximately $3.5~\mathrm{K}$, a temperature response comparable to that induced by a doubling of the atmospheric CO\textsubscript{2} concentration~\citep{IPCC_2021_WGI_Ch_7}.
This result is remarkable: despite no external changes in the radiative properties of the atmosphere and without injecting any heat into the system, we nevertheless simulate a strong surface warming that can only be explained by the radiative response of the atmosphere to the imposed circulation change.
These radiative forcings and feedbacks are explored in Section~\ref{sec:forcing/feedback}.

\subsection{Precipitation} \label{sec:precip_changes}

\begin{figure}[t]
    \centering
  \noindent\includegraphics[width=39pc,angle=0]{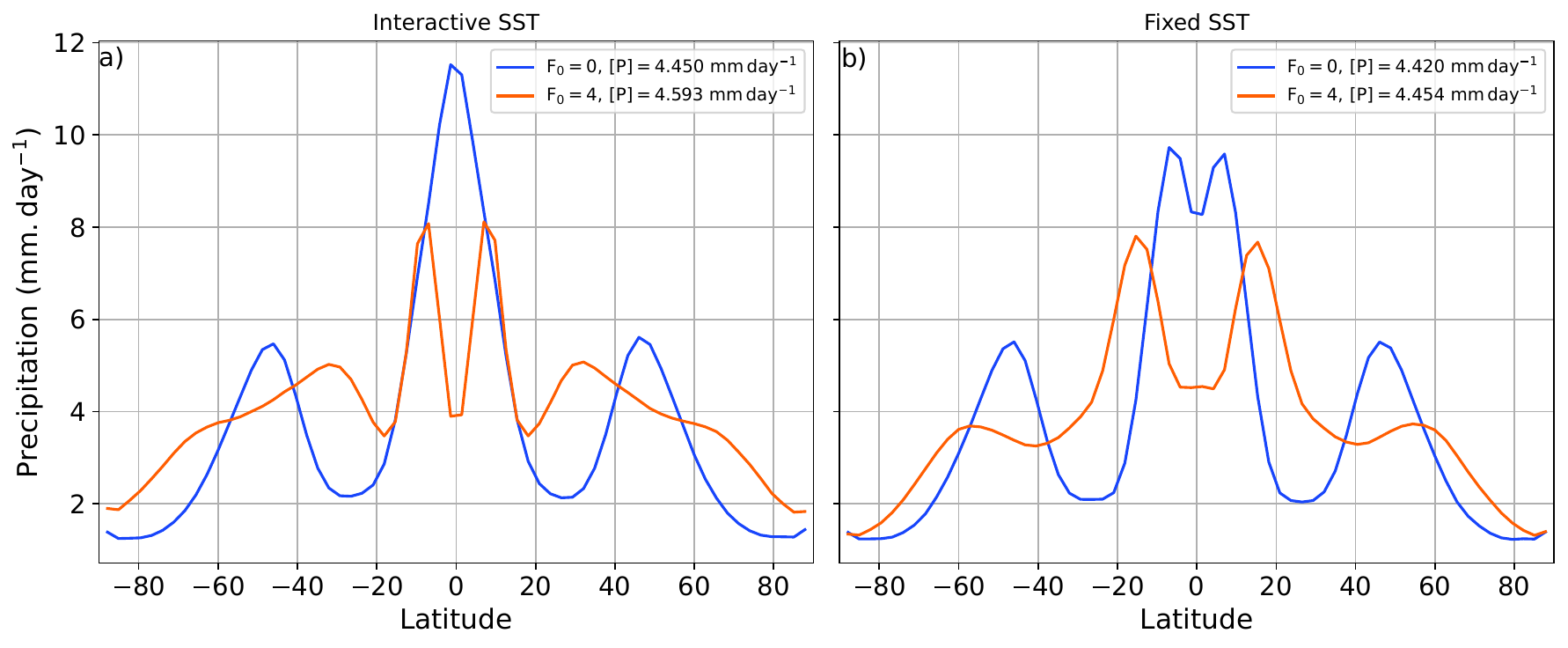}\\
  \caption{Zonally- and time-averaged precipitation rates for two different values of $F_0$: $F_0=0$~m.s\textsuperscript{-1}.day\textsuperscript{-1} (control run, blue curve) and $F_0=4$~m.s\textsuperscript{-1}.day\textsuperscript{-1} (superrotating state, orange curve), in the interactive-SST (\textbf{a}) and fixed-SST experiments (\textbf{b}). The value of the globally-averaged precipitation rate in each case is given in the legend.}\label{fig:precip_profile}
\end{figure}

The zonally- and time-averaged precipitation rates and the globally-averaged precipitation rates for the different values of $F_0$ and different experiment setups are shown in Figure~\ref{fig:precip_profile}.
In the interactive-SST control run (Fig.~\ref{fig:precip_profile}, \textbf{a}), the precipitation rate exhibits the standard meridional structure with a dominant peak located at the equator (i.e., the ITCZ) and two minima in subtropical regions, centered around $30^\circ$, coinciding with the descending branches of the Hadley cells (Fig.~\ref{fig:jet}). There are two secondary peaks in the mid-latitudes, at approximately $45^{\circ}$, on the poleward flanks of the jet stream; poleward of $45^{\circ}$ the precipitation rate decreases sharply.
The control run in the fixed-SST case is very similar (Fig.~\ref{fig:precip_profile}, \textbf{b}), with the only exception being that there is a split ITCZ with two roughly symmetric peaks located approximately $5^\circ$ north and south of the equator.
The globally-averaged precipitation rates for the interactive- and fixed-SST cases are similar.

In the superrotating state, the globally-averaged precipitation rate increases slightly in the interactive-SST case (Fig.~\ref{fig:precip_profile}, \textbf{a}), but less than the $2-3$~\%.K\textsuperscript{$-1$} usually found in global warming experiments~\citep{ogormanEnergeticConstraintsPrecipitation2012}.
In the fixed-SST case, the increase in globally-averaged precipitation rate is even weaker: these two results are in agreement with tropospheric energy budget arguments, based on the approximate balance between radiative cooling and latent heat release above cloud base~\citep{ogormanEnergeticConstraintsPrecipitation2012}.
Perhaps more interestingly, we find a drastic change in the meridional distribution of precipitation in the superrotating state: Precipitation is strongly decreased in the tropics, slightly decreased in the mid-latitudes (the two regions where the precipitation rate has local maxima in the control climate) and increased in the subtropics and at high latitudes, leading overall to a much more homogeneous spatial distribution of precipitation compared to the current climate.
These changes are qualitatively similar between the interactive- and fixed-SST experiments, although there are some differences: In the superrotating states, the emergent precipitation maxima in the subtropics are further away from the equator in the fixed-SST runs, and there are no maxima of precipitation at $30^{\circ}$ latitude as in the interactive-SST runs.
These changes in the precipitation profiles are consistent with the circulation changes described above.
Indeed in the current climate, the Hadley cells transport moisture from the subtropics towards the equator~\citep{hartmannGlobalPhysicalClimatology2016}.
The weakening of this overturning circulation, shown in Figure~\ref{fig:jet}, provides a possible explanation for the changes in tropical precipitation.
The processes related to changes in the precipitation distribution are discussed in more detail in Section~\ref{sec:p-e}

\section{From circulation changes to energy budget changes}\label{sec:forcing/feedback}

In order to link  the climate changes described in the previous section to the imposed circulation change, we will investigate how these circulation changes influence the distribution of energy in the climate system.
In particular, we discuss the roles of the mean meridional circulation and of the eddies in Section~\ref{sec:forcing/feedback}.\ref{sec:energy_transport_changes}, and then use the methods introduced in Section~\ref{sec:model} to discuss the radiative budget changes in Section~\ref{sec:forcing/feedback}.\ref{sec:radiative_kernel_rh}.

\subsection{Meridional energy transport}\label{sec:energy_transport_changes}

We first compute the changes in vertically-integrated meridional energy transport related to the transition to superrotation, and connect these changes to the atmospheric circulation changes described in Section~\ref{sec:changes}.\ref{sec:circulation_changes}.
We denote these changes as $\delta  \left( \overline{\nabla \cdot \left[ \mathbf{v} \mathrm{MSE} \right]} \right)$, where $\mathbf{v}$ is the horizontal velocity vector and $\mathrm{MSE}$ is the moist static energy, i.e. $ \mathrm{MSE}~=~c_p T~+~g z~+~L q$, where $c_p$ is the specific heat capacity of air at constant pressure and $L$ is the specific latent heat of vaporization (the other symbols have their usual meanings).
The overline represents the time- and zonal-average and the brackets represent the vertical mass-weighted integral.

\begin{figure}
  \centering
  \includegraphics[width=36pc]{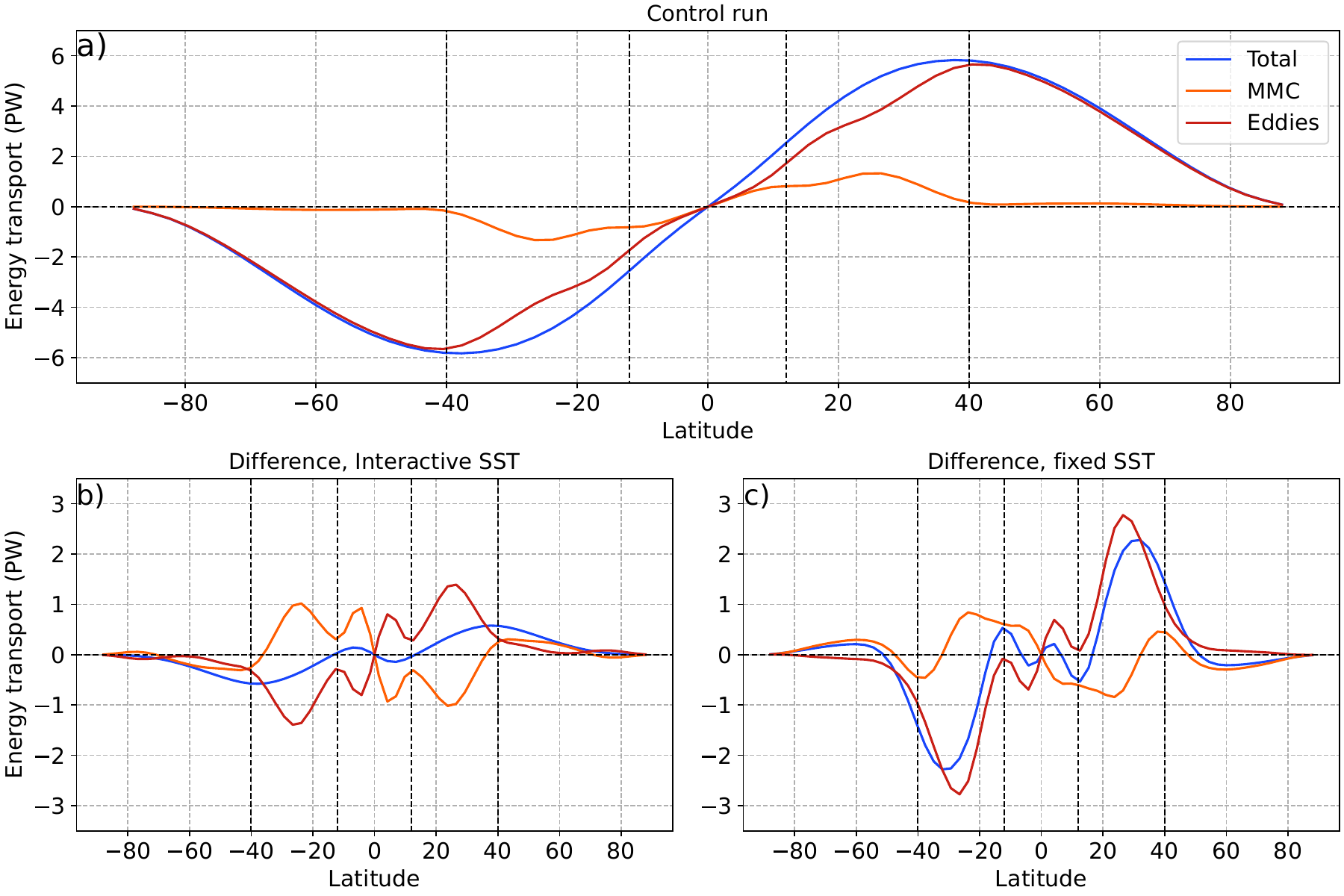}
  \caption{\textbf{a)} total atmospheric meridional energy transport, positive northward, along with its mean meridional circulation and eddy components in the control run with interactive SST.
  Bottom: Transport changes between the superrotating ($F_0 = 4$~m.s\textsuperscript{$-1$}.day\textsuperscript{$-1$}) and control ($F_0=0$~m.s\textsuperscript{$-1$}.day\textsuperscript{$-1$}) runs, for interactive (\textbf{b}) and fixed SST (\textbf{c}).}
  \label{fig:MSEtransport}
\end{figure}

After applying a correction to the meridional wind in order to compensate the numerical errors in the closure of the energy balance~\citep{bangalathNewMassFlux2020}, we compute explicitly the meridional MSE flux and decompose it into mean meridional circulation (MMC) and eddy components (Fig.~\ref{fig:MSEtransport}).
Here, the eddies include both transient and standing eddies, but since the boundary conditions are axisymmetric in our problem, standing eddies are not expected to play a very important part.

\begin{figure}
    \centering
    \includegraphics[width=36pc]{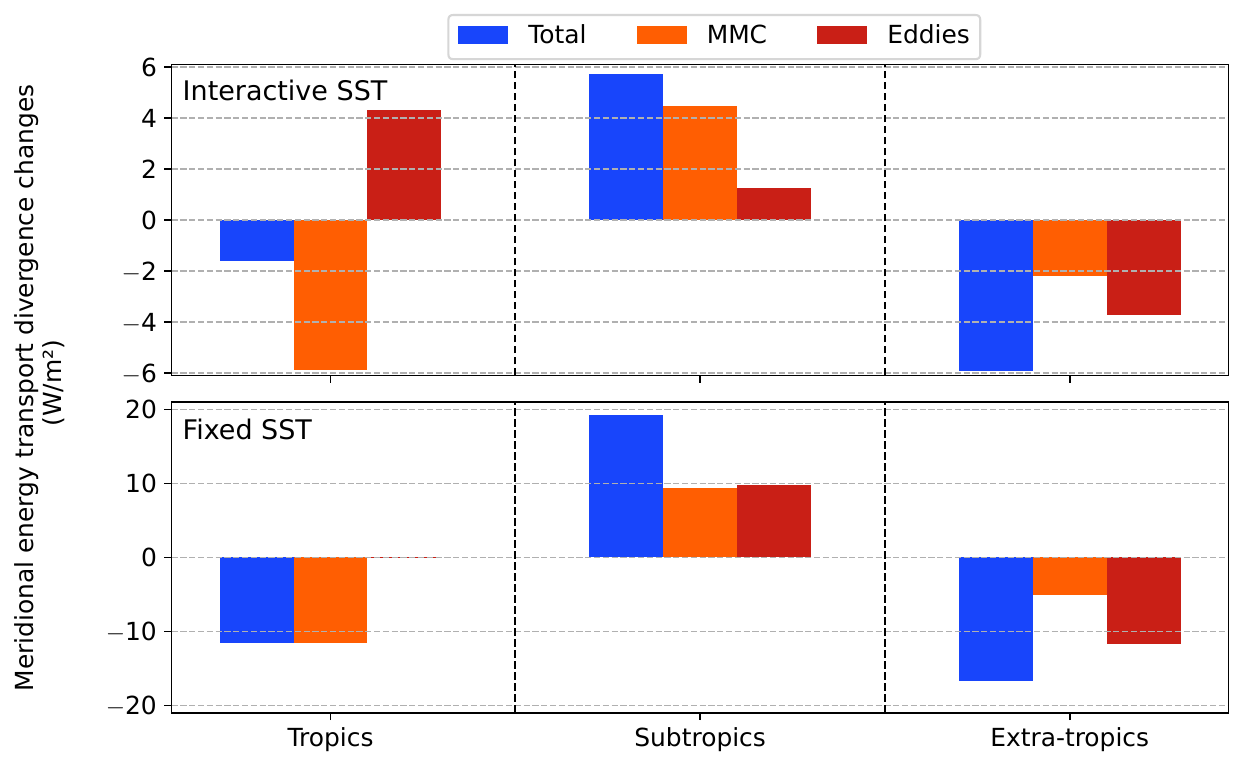}
    \caption{Changes in meridional energy transport divergence between the superrotating run ($F_0=4$~m.s\textsuperscript{$-1$}.day\textsuperscript{$-1$}) and the control run ($F_0 = 0$~m.s\textsuperscript{$-1$}.day\textsuperscript{$-1$}), averaged over three regions: tropics ($12^{\circ}$S to $12^{\circ}$N), subtropics ($40^{\circ}$S to $12^{\circ}$S and $12^{\circ}$N to $40^{\circ}$N), and mid-/high-latitudes ($90^{\circ}$S to $40^{\circ}$S and $40^{\circ}$N to $90^{\circ}$N), for interactive- (top panel) and fixed-SST runs (bottom panel). The total (blue bars), mean meridional circulation (orange bars) and eddies (red bars) components of these changes are shown.  }
    \label{fig:MSEtransport_groupedbar}
\end{figure}

We first note that the energy transport in the control run (Fig.~\ref{fig:MSEtransport}, panel \textbf{a}) exhibits the standard characteristic features: both the MMC and transient eddies transport energy poleward, the latter dominating everywhere except at low latitudes.
The meridional energy flux divergence for the MMC is positive below $25^\circ$ in the ascending branch of the Hadley cell, and negative between $25^\circ$ and $40^\circ$ in the descending branches.
For the eddy component, divergence is positive below $40^\circ$ and negative on the poleward flank of the jet.

In the superrotating state (Fig.~\ref{fig:MSEtransport}, panels \textbf{b} and \textbf{c}), the most evident change is that the poleward eddy energy flux (green curve) increases at all latitudes, while the magnitude of the energy flux associated with the MMC (orange curve) mostly decreases in the tropics and subtropics, and becomes weakly positive in the mid-latitudes.
The former is broadly consistent with the appearance of a turbulent jet at the equator, while the latter can be explained qualitatively by the collapse of the Hadley cells at low latitudes and the poleward shift of the Ferrell cells at high latitudes (the two regions are separated by a buffer region where energy transport changes due to the MMC are small, around 40$^\circ$).

The partial compensation of these opposing effects determines the pattern of total poleward energy transport change (blue curve): in the interactive SST case, it is negative in the tropics (roughly until $12^{\circ}$ latitude), with a minimum near $10^{\circ}$, and positive at higher latitudes with a maximum near $40^{\circ}$ latitude.
This indicates that the MMC mostly drives the total energy transport changes in the tropics and high latitudes, and the eddies do in the subtropics.
As a result, the change in net MSE flux divergence is negative in the tropics and mid-latitudes, and positive in the subtropics (Fig.~\ref{fig:MSEtransport_groupedbar}).

While the general conclusions in the above paragraph apply qualitatively to the fixed-SST and interactive-SST runs, the former exhibit a number of specific features.
To start, there is a small region around the equator (roughly $5^\circ$ north and south) where the effect of eddies dominates the transport changes, resulting in increased poleward transport compared to the control run, but this is more than compensated for by the changes in the rest of the tropics and the net energy transport into the whole region (approximately $0^\circ$-$10^\circ$) remains larger than in the control run.
More generally, the net transport changes follows more closely the eddy component, due to the smaller contribution of the MMC at low latitudes (compared to the interactive-SST case) associated with less drastic changes in the structure of the overturning cells.
This is particularly the case in the subtropics, where the maximum in eddy transport changes (at approximately 30$^\circ$) is much stronger than in the interactive-SST runs.
As a result, the net energy convergence integrated over the subtropical and mid-latitude regions is much larger too (Fig.~\ref{fig:MSEtransport_groupedbar}).

Below, we relate the energy transport changes described above to radiative budget changes, which can in turn be interpreted in terms of changes in physical fields such as temperature and atmospheric moisture content.

\subsection{Radiative budget changes and their relation to relative humidity} \label{sec:radiative_kernel_rh}

Writing the zonally- and time-averaged energy balance at a given latitude enables us to link TOA radiative fluxes $R$ (positive downwards) to the meridional energy flux introduced above.
The energy balance for a volume of air with infinitesimal meridional extent can be written as: $R\equiv \overline{F}_{\mathrm{SW}}-\overline{\mathrm{OLR}} = \overline{\nabla \cdot \left[ \mathbf{v} \mathrm{MSE} \right]} - \overline{F_{\mathrm{sfc}}}$, where $F_{\mathrm{SW}}$ is the net incoming shortwave radiation, $\mathrm{OLR}$ is the outgoing longwave radiation, and $F_{\mathrm{sfc}}$ is the net upward surface energy flux. We do not consider shortwave flux changes here, as they empirically prove to be negligible.
It follows that

\begin{equation}
  \delta \overline{R} = - \delta \left( \overline{\mathrm{OLR}} \right) = \delta  \left( \overline{\nabla \cdot \left[ \mathbf{v} \mathrm{MSE} \right]} \right)-\delta(\overline{F_{\mathrm{sfc}}}).
\end{equation}
For the interactive-SST runs, surface energy balance imposes that the changes in surface flux vanish, $\delta(\overline{F_{\mathrm{sfc}}})=0$, because they must always equilibrate to the prescribed ocean heat transport; in this case, the TOA radiative changes exactly balance the meridional energy flux convergence changes.
However, this need not be the case in the fixed-SST runs where the surface fluxes are not constrained by energy balance.

\begin{figure}[t]
    \centering
  \noindent\includegraphics[width=36pc,angle=0]{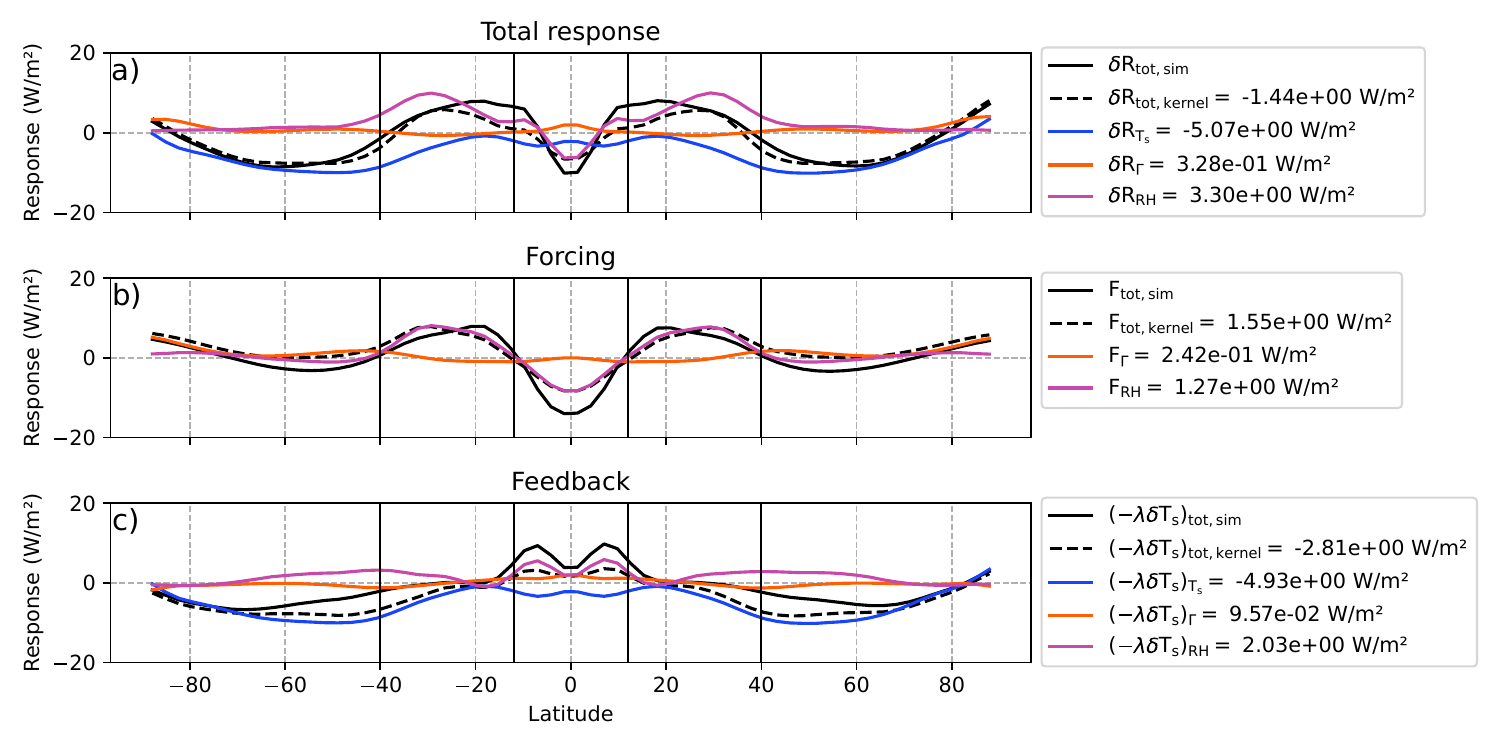}\\
  \caption{Total response and the different components of the decomposition of TOA radiative anomalies using relative humidity, computed using radiative kernels, between the $F_0 = 0$ and $F_0 = 4$~m.s\textsuperscript{$-1$}.day\textsuperscript{$-1$} experiments.
  Panel \textbf{a} corresponds to the interactive-SST runs, panel \textbf{b} to the fixed-SST runs, and panel \textbf{c} is the difference between the two.
  Numerical values given in the legend correspond to the global average of the associated component.}\label{fig:decomp_rh}
\end{figure}

We show in Figure~\ref{fig:decomp_rh} the TOA radiative budget changes (black line) in the interactive-SST runs (panel \textbf{a}), in the fixed-SST runs (the radiative forcing, panel \textbf{b}) and the difference between the two (i.e., the feedbacks, panel \textbf{c}).
The meridional structure of the TOA radiative changes in the interactive-SST runs is, as expected, in agreement with the analysis of the energy transport changes: the excess of energy transport in the deep tropics (between $10^\circ$S and $10^\circ$N) and in the mid- and high-latitudes (poleward of $40^\circ$) is balanced by increased OLR, while the opposite holds in a broad subtropical region (between $10^\circ$ and $40^\circ$).
The net TOA radiative budget changes (panel \textbf{a}) are dominated by the radiative forcing (panel \textbf{b}) in the low-latitudes (between $40^\circ$S and $40^\circ$N) and by radiative feedbacks at higher latitudes (poleward of $40^\circ$).

To relate the changes in TOA radiative budget, in particular the radiative forcing and feedback components, to physical fields, we use radiative kernels (as described in Section~\ref{sec:model}\ref{sec:radiative_kernel}).
We choose the relative humidity-based decomposition introduced by~\citet{Held2012a} and~\citet{Ingram2012} [see also~\cite{Jeevanjee2021}], in which relative humidity is held fixed in the Planck and lapse-rate terms.
In addition to the general arguments for this method, including that it avoids unphysical scenarios (e.g. cooling at fixed specific humidity leading to supersaturation) and disentangles the humidity and lapse rate effects related to the Clausius-Clapeyron relation, in our particular case it provides a much simpler interpretation of the radiative budget changes, as discussed in Appendix~A.
This relative humidity-based decomposition results in contributions from changes in relative humidity (Fig.~\ref{fig:decomp_rh}, pink curves), lapse rate (orange curves), and surface temperature (Planck response, blue curve) to changes in TOA radiation.
Similar decompositions apply for radiative forcing and feedbacks: $F_{\mathrm{tot, kernel}}=F_{\Gamma}+F_{\mathrm{RH}}$ and $-(\lambda\delta T_{\mathrm{s}})_{\mathrm{tot, kernel}}=-{(\lambda\delta T_{\mathrm{s}})}_{\mathrm{T_{\mathrm{s}}}}-{(\lambda \delta T_{\mathrm{s}})}_{\Gamma}-{(\lambda\delta T_{\mathrm{s}})}_{\mathrm{RH}}$; these components are represented in Fig.~\ref{fig:decomp_rh} with the same color code.
All these terms can be expressed using the original kernel computed in~\citet{liuRadiativeKernelsIsca2020}, as explained in Appendix B.

First, we compare the total radiative budget changes computed using the kernels (Fig.~\ref{fig:decomp_rh}, dashed black curves) with the changes computed directly from the model outputs (Fig.~\ref{fig:decomp_rh}, solid black curves).
Overall, the meridional structures of the two curves are very similar, though there are differences due both to the linearization inherent to the kernel method and to the fact that we use kernels constructed with a different radiation scheme.
In particular, the domination of the radiative forcing and feedback contributions at low (below $40^{\circ}$) and mid-/high-latitudes (above $40^{\circ}$) respectively is even more pronounced in the radiative kernel budget.
In the following, we will embrace this approximation and explain qualitatively why the low latitudes are dominated by the forcing and the mid- and high-latitudes by the feedbacks.
We note that the smaller contributions of the radiative forcing at mid- and high-latitudes is at odds with the larger energy transport changes reported in those regions in the fixed-SST runs in Section~\ref{sec:forcing/feedback}.\ref{sec:energy_transport_changes}.
This can be explained by the fact that the surface fluxes, as noted above, are not constrained in this case, and indeed their changes absorb most of the energy transport changes, leaving the TOA radiative budget largely unaltered.

\begin{figure}
    \centering
    \includegraphics[width=36pc]{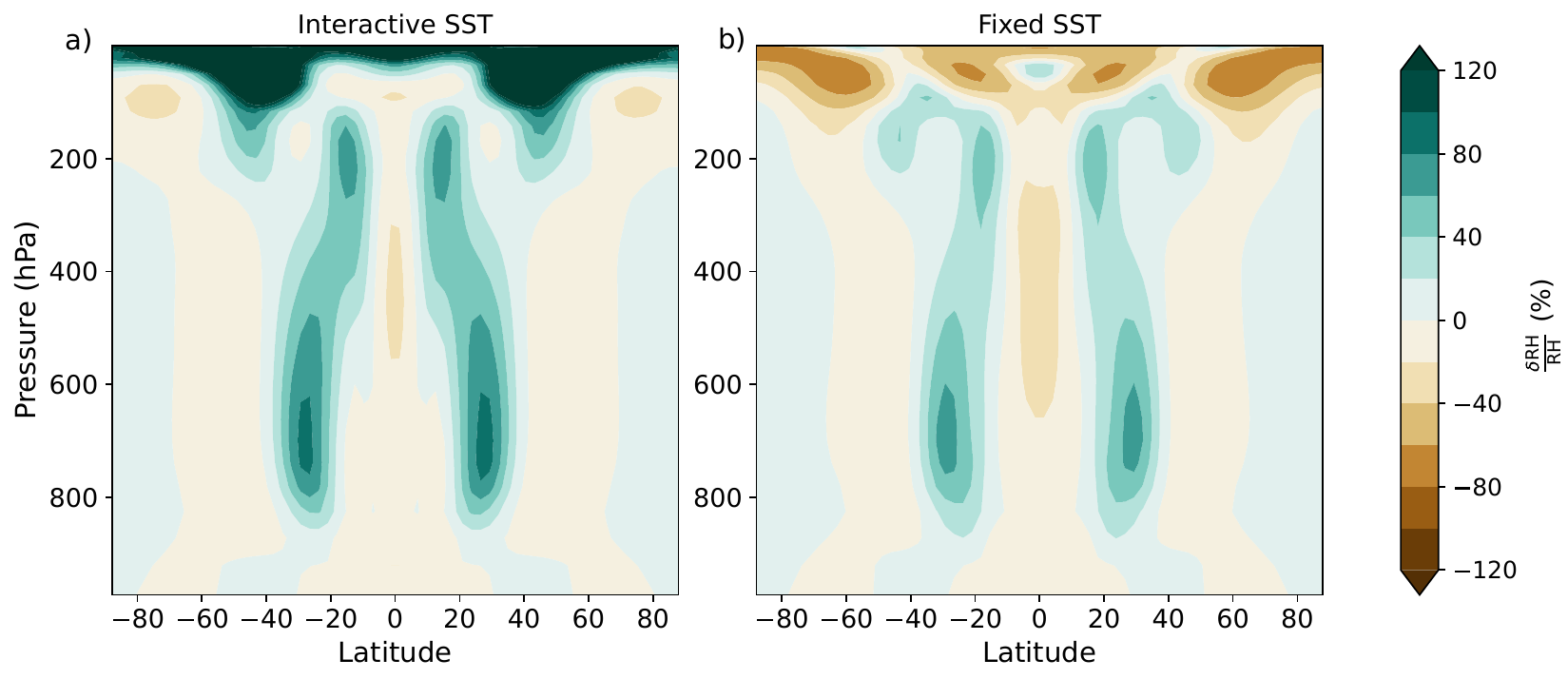}
    \caption{Time- and zonally-averaged fractional relative humidity changes between the $F_0~=~4$~m.s\textsuperscript{$-1$}.day\textsuperscript{$-1$} and $F_0~=~0$~m.s\textsuperscript{$-1$}.day\textsuperscript{$-1$} experiments, for the interactive-SST (panel \textbf{a}) and fixed-SST (panel \textbf{b}) experiments.}    \label{fig:rh_changes}
\end{figure}

We now consider the decomposition of the TOA radiative flux changes in the interactive-SST runs (panel \textbf{a}), in the fixed-SST runs (i.e., the radiative forcing, panel \textbf{b}) and the difference (i.e., the feedbacks, panel \textbf{c}) into their lapse rate, relative humidity and Planck components (Figure~\ref{fig:decomp_rh}, color curves).
The results are strikingly clear: the lapse rate contributions are vanishingly small for both the radiative forcing and feedbacks (and, therefore, for the net radiative changes also), and the radiative forcing (panel \textbf{b}) is almost entirely explained by the relative humidity changes (red curve), which is confined to the tropics and subtropics and largely vanishes elsewhere.
We can therefore compare the meridional structure of the radiative forcing to the relative humidity changes, shown in Figure~\ref{fig:rh_changes} (panel \textbf{b} for the fixed-SST runs).

Indeed, the OLR increase in the tropics (within $12^\circ$ latitude of the equator) corresponds to a decrease in relative humidity, while the OLR decrease in the subtropics and until $40^\circ$ is associated with an increase in relative humidity.
Note that although the relative humidity changes are stronger in the subtropics, the two effects lead to radiative forcings of similar magnitudes (but opposite in sign) because the radiative kernel is larger closer to the equator.
For the same reason, the relative humidity changes poleward of $40^\circ$ do not lead to appreciable radiative forcing.

A second notable finding is that the relative humidity feedback is also very small; indeed, the relative humidity changes in the interactive-SST runs (Fig.~\ref{fig:rh_changes}, panel \textbf{a}) are very similar to those for the fixed-SST runs.
As a consequence, the feedback becomes strongly dominated by the Planck response.
It is also very close to zero in the tropics and subtropics and has a substantial contribution only poleward of $40^\circ$.

\subsection{Summary}

We conclude the following about the connection between the circulation changes and the TOA radiation changes:

\begin{itemize}
\item In the deep tropics, between $0^\circ$ and $12^{\circ}$ latitude, the radiative changes are mainly driven by the radiative forcing (Fig.~\ref{fig:decomp_rh}). The circulation changes (specifically the collapse of the MMC, which is not fully compensated by eddy transport changes) result in decreased divergence of the meridional energy transport (Fig.~\ref{fig:MSEtransport_groupedbar}), which is balanced by increased OLR. The increased OLR, in turn, is driven by a drying of the atmosphere (Fig.~\ref{fig:rh_changes}) which again finds its source in the circulation changes, as they result in increased divergence of the meridional moisture transport (see Section~\ref{sec:p-e}). This behavior is found in both fixed- and interactive-SST runs, although in the latter case a portion of the additional energy convergence can be used to warm the surface and the whole column, increasing its specific humidity at the same time (see Appendix~A). This feedback occurs essentially at fixed relative humidity.

\item Further from the equator, in the subtropics (between $12^\circ$ and $40^{\circ}$ latitude), the circulation changes (combination of reduced mean meridional circulation and increased eddy transport) lead to decreased convergence of meridional energy transport (Fig.~\ref{fig:MSEtransport_groupedbar}), balanced by decreased OLR. The OLR decrease mostly results from a strong increase in relative humidity (Fig.~\ref{fig:rh_changes}). Again, this relative humidity change is consistent with the circulation changes, and in particular the collapse of the MMC. The interactive-SST runs are quite similar to the fixed-SST runs in that area, i.e. the feedbacks remain relatively small, as the deficit of energy convergence can be compensated radiatively through the humidity increase without affecting surface temperature. Nevertheless, this is less and less the case as we move poleward, and there is a progressive transition to the regime described next.

\item Poleward of $40^{\circ}$, the circulation changes lead to increased convergence of meridional energy transport, driven by a combination of the MMC and the eddy transport (Fig.~\ref{fig:MSEtransport_groupedbar}), which provide contributions of comparable magnitudes. Like in the tropics, this is balanced by increased OLR, but the adjustment mechanism is quite different. Although the relative humidity decreases in the equatorward part of the region and increases in the poleward part, the associated radiative response is very small, similar to the lapse rate response. Consequently, the balance is reached by uniformly warming the column and the surface, i.e. by the Planck response (because this is not possible in the fixed-SST case, the surplus of energy is absorbed by the surface fluxes).

\end{itemize}

\section{Linking changes in the atmospheric energy budget and surface temperature: a moist static energy diffusion model}\label{sec:ebm}

The above analysis suggests a simple picture, whereby circulation changes directly affect the surface temperature in mid-latitudes through meridional energy transport changes, and affect it in the tropics through the radiative effects of relative humidity changes, themselves induced by the forced circulation change.
To further support this argument, we relate the surface temperature changes described in Section~\ref{sec:changes} to the radiative budget changes described in Section~\ref{sec:forcing/feedback} through an energy balance model (EBM).
This class of models solves the space-dependent energy balance equations for temperature, without representing explicitly the circulation of the atmosphere, by assuming that the horizontal energy transport can be viewed as a diffusive process:

\begin{equation} \label{eq:diffusion}
    C \partial_t T_{\mathrm{s}} = R - G + \nabla \cdot \left( D \nabla h \right),
\end{equation}
where $C$ is the heat capacity of the layer where the diffusion process happens, $R$ is the TOA radiative budget (positive downwards), $G$ is the surface radiative budget (positive downwards), $D$ is the diffusivity coefficient and $h$ the generalized enthalpy used in the diffusive closure.
Historically, dry static energy was used for the diffusive closure, and the model resulted in a diffusion equation for temperature, with sources and sinks determined by the radiative budget, providing perhaps the simplest climate model with spatial structure~\citep{Budyko1969,Sellers1969,North1981}.
More recently it has been argued that such models could also describe finer scale properties of climate, such as the regional response to radiative forcing, for instance in the context of polar amplification, provided the water vapor feedback is taken into account by using MSE as the diffusing quantity~\citep{Flannery1984, Rose2014, Merlis2018}.

\begin{figure}
    \centering
    \includegraphics[width=36pc]{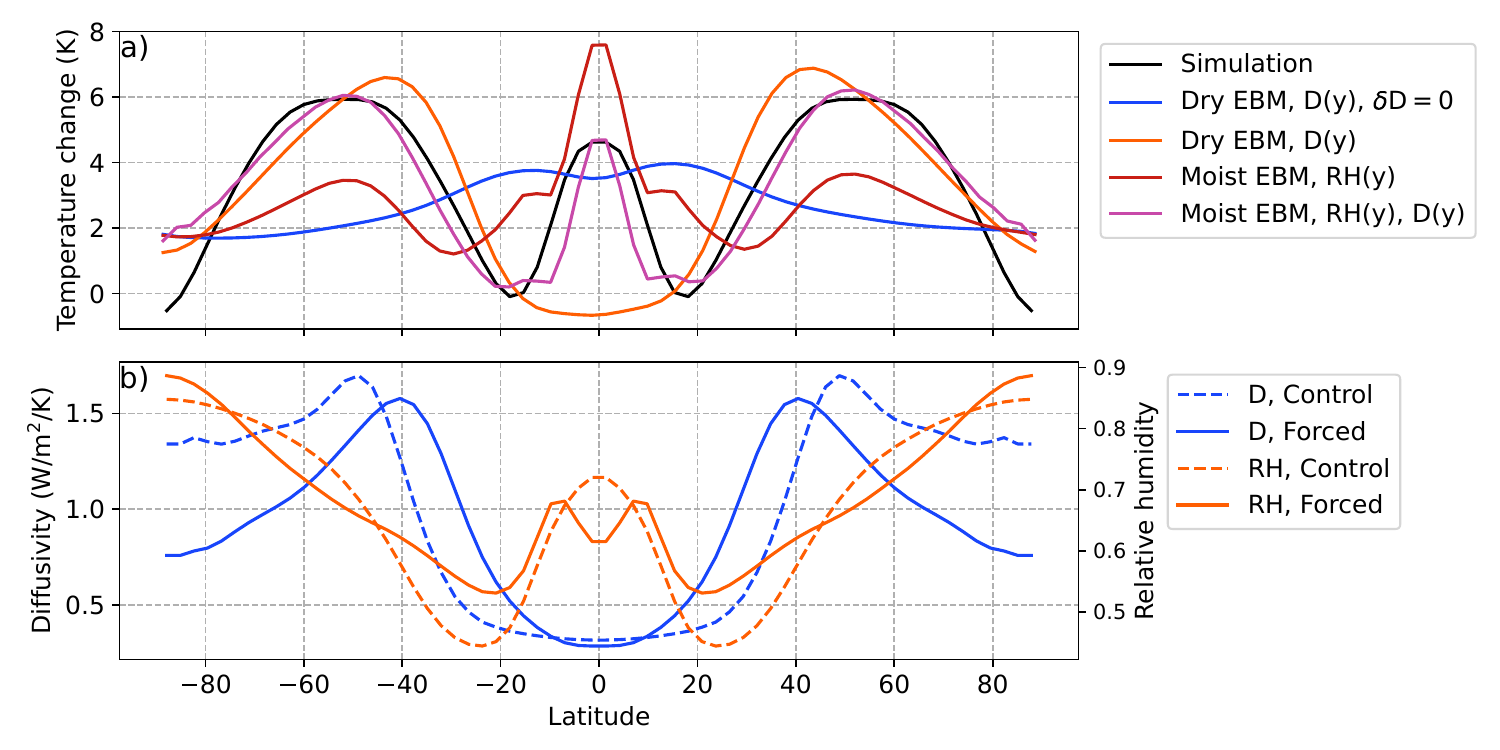}
    \caption{\textbf{a)} Near-surface temperature difference between the superrotating and control GCM runs (black line), and for the different formulations of the energy balance model: dry model without (blue) and with diffusivity changes (orange) and moist model with relative humidity changes, without (green) and with (pink) diffusivity changes.
    \textbf{b)} Diffusivity (blue) and near-surface relative humidity (orange) profiles for the control climate (dashed lines) and superrotating climate (solid lines).}
    \label{fig:EBM}
\end{figure}

Here we use this moist EBM framework as a way to relate energy budget changes, due to meridional energy transport and radiative budget changes, to surface temperature.
In particular, we seek to explain how comparable energy budget changes in the tropics, subtropics, and extratropics result in small, negligible, and large surface temperature changes, respectively. To do so, we solve the 1D diffusion equation for the control and forced runs by prescribing the meridional structure of the radiative budget from the zonally-averaged GCM simulation output, and plot the surface temperature difference on Figure~\ref{fig:EBM} (panel \textbf{a}), for four different levels of complexity (below we denote meridional derivatives with $\partial_{y}$):

\begin{itemize}
\item \emph{Dry model with constant diffusivity:} Temperature profile changes at equilibrium satisfy $\delta R+\partial_y \left(D \partial_y \delta T\right)=0$ [see (\ref{eq:diffusion})]. We use either a uniform diffusivity or a latitude-dependent diffusivity computed empirically from the control run so that the relation
  \begin{equation}
    \overline{\left[ v \cdot \mathrm{MSE} \right]}=D\partial_{y}\overline{\mathrm{MSE}_{s}}\label{eq:diffusivity}
  \end{equation}
  is satisfied (with a regularization at the equator since both the meridional MSE transport and the meridional MSE gradient vanish).
Here, the brackets denote vertical integration, the overline denotes a time and zonal average and the subscript ``s'' denotes surface quantities.
The expression for the diffusivity coefficient comes from the hypothesis that the meridional energy flux is proportional to the gradient of the chosen generalized enthalpy at the surface (i.e. moist static energy in this case).
In both cases, we use the same diffusivity for the control and superrotating runs.
Taking into account, or not, the meridional structure in diffusivity does not affect the results.

\item \emph{Dry model with varying diffusivity:} Temperature profile changes satisfy $\delta R+\partial_y \delta(D \partial_y T)=0$, with $D$ computed from (\ref{eq:diffusivity}) separately for the two states; the two profiles are represented in Figure~\ref{fig:EBM} (panel \textbf{b}). We note that the structure of diffusivity changes significantly, in agreement with the circulation changes described in Section~\ref{sec:circulation_changes}: In the control run the diffusivity is particularly strong on the poleward flank of the mid-latitude jets where baroclinic eddies transport heat efficiently. But in the superrotating state these jets move poleward and weaken, resulting in reduced diffusivity. On the other hand, the formation of the equatorial jet and the associated instabilities enhance lateral mixing at low-latitudes compared to the control run.

\item \emph{Moist model with varying relative humidity:} Temperature changes satisfy $\delta R+\partial_y \left( D \partial_y \left( \delta \mathrm{MSE}/c_p\right) \right)=0$, where $\delta \mathrm{MSE}=c_p\delta T+L_{\rm v}\delta(\mathrm{RH} q^{\star}(T))$. The relative humidity profiles for the two states are shown in Figure~\ref{fig:EBM} (panel \textbf{b}).

\item \emph{Moist model with varying relative humidity and diffusivity:} Temperature changes satisfy $\delta R+\partial_y \delta \left(D \partial_y \mathrm{MSE}/c_p\right)=0$.
\end{itemize}

We use the \textit{dedalus} framework~\citep{dedalus} to solve for the equilibrium (i.e., steady state) temperature distribution as a boundary value problem, where we constrain the global average of temperature to be the same as in the GCM simulation and assume the meridional flux of energy vanishes at the poles.
In the moist case, for the sake of simplicity, we linearize the temperature dependence of the saturation specific humidity, writing $q^{\star}(T) = \alpha T + \beta$.

The simplest case of a dry model with constant diffusion acting on a meridional temperature profile that would be consistent with the radiative budget changes fails dramatically at explaining the GCM-simulated temperature response (Fig.~\ref{fig:EBM}, panel \textbf{a}). In this case the concavity of the surface temperature change has opposite sign to the radiative budget changes, which leads to the existence of a local maximum in the subtropics and local minimum at the equator.
Hence, the dry model does not capture the simulated strong extratropical warming with a local maximum in mid-latitudes, the local minimum in the subtropics and the local maximum at the equator (although the value of the predicted surface temperature change there matches the GCM-simulated change).

Our second level of complexity takes into account circulation changes in a simple way by allowing the diffusivity to vary between the control and the superrotating states. The resulting surface temperature changes (Fig.~\ref{fig:EBM}, panel \textbf{a})  capture the warming of the extratropics with the correct magnitude.
However, this modified dry model still fails to replicate warming in the deep tropics; in a dry diffusive picture, the prescribed increased OLR can only be compensated by reducing the meridional temperature gradient in this region.
On the other hand, in a moist model the meridional MSE gradient can be reduced while still increasing the meridional temperature gradient if the meridional gradient of relative humidity is reduced concurrently.
The forcing/feedback analysis above suggests that this happens as a consequence of the collapse of the meridional overturning circulation.

Figure~\ref{fig:EBM} (panel \textbf{a}, green curve) shows the surface temperature predicted by the EBM for consistent radiative budget changes and relative humidity changes, but without any diffusivity changes. We note that this moist model predicts warming at the equator but of a magnitude that is much stronger than that observed in the GCM simulations. If we also take into account diffusivity changes resulting from the presence of the strong equatorial jet, the EBM forced with the GCM-simulated radiative budget changes predicts surface temperature changes (Fig.~\ref{fig:EBM}, panel \textbf{a}) that are very similar to those found in the GCM simulations, showing that diffusivity changes play a mitigating role in the tropics compared to the changes in the meridional distribution of relative humidity alone.

Overall, solving this simple energy balance model in different configurations revealed the ingredients needed to reproduce the simulated surface temperature changes in two distinct regions.
The relative humidity changes induced by the imposed circulation changes are key to reproduce the tropical warming, whereas the mid-latitude response is driven by changes in the way the circulation transports energy.

\section{Moisture transport and precipitation changes}
\label{sec:p-e}

As discussed in Section~\ref{sec:changes}, the precipitation profile suffers an important structural change when a superrotating jet is forced at the equator.
The goal of this section is to understand how these changes are related to the circulation changes.
Specifically, we will study changes in moisture transport, which are related to the rates of precipitation $\mathrm{P}$ and evaporation $\mathrm{E}$ through the water mass budget over an atmospheric column (between the surface and the top of the atmosphere): $\mathrm{P-E} = -\nabla \cdot \left[ \overline{\mathbf{u}q}\right]$.
Because evaporation changes are typical small, the moisture convergence ($\mathrm{P-E}$) changes are essentially determined by the precipitation changes.

It follows from the water mass budget that the $\mathrm{P-E}$ changes associated with a change in $F_0$ can be decomposed into a sum of different contributions, as done for example by \citet{seagerThermodynamicDynamicMechanisms2010, byrneResponsePrecipitationMinus2015} for anthropogenic climate change:

\begin{equation}
\label{decomposition}
    \begin{split}
        \delta \left( \mathrm{P-E} \right) & = -\nabla \cdot \left[ \delta \left( \overline{\mathbf{u}q} \right) \right], \\
         & = \underbrace{-\nabla \cdot \left[ \overline{\mathbf{u}} \delta \left( \overline{q} \right) \right]}_{\delta\mathrm{MTh}} \underbrace{-\nabla \cdot \left[ \delta \left(\overline{\mathbf{u}} \right) \overline{q} \right]}_{\delta \mathrm{MDyn}}  \underbrace{-\nabla \cdot \delta \left[ \overline{\mathbf{u}^{\prime} q^{\prime}} \right]}_{\delta \mathrm{Eddy}} \underbrace{-\nabla \cdot \left[ \delta \left( \overline{\mathbf{u}} \right) \delta \left( \overline{q} \right) \right]}_{\delta \mathrm{NL}}.
    \end{split}
\end{equation}
In Eq. (\ref{decomposition}), each term on the right hand side corresponds to a different physical process that can cause a change in $\mathrm{P-E}$:

\begin{itemize}
    \item $\delta \mathrm{MTh}$ (which stands for "Mean Thermodynamic") contains the contribution due to changes in the zonally- and time-averaged specific humidity distribution with fixed MMC;
    \item $\delta \mathrm{MDyn}$ (which stands for "Mean Dynamic") contains the contribution due to changes in the MMC with fixed mean specific humidity distribution;
    \item $\delta \mathrm{Eddy}$ contains the contribution due to changes in transient eddy moisture fluxes;
    \item $\delta \mathrm{NL}$ contains the contribution due to non-linear effects associated with correlations between changes in the MMC and in the mean specific humidity distribution.
\end{itemize}

\begin{figure}
    \centering
    \includegraphics[width=36pc]{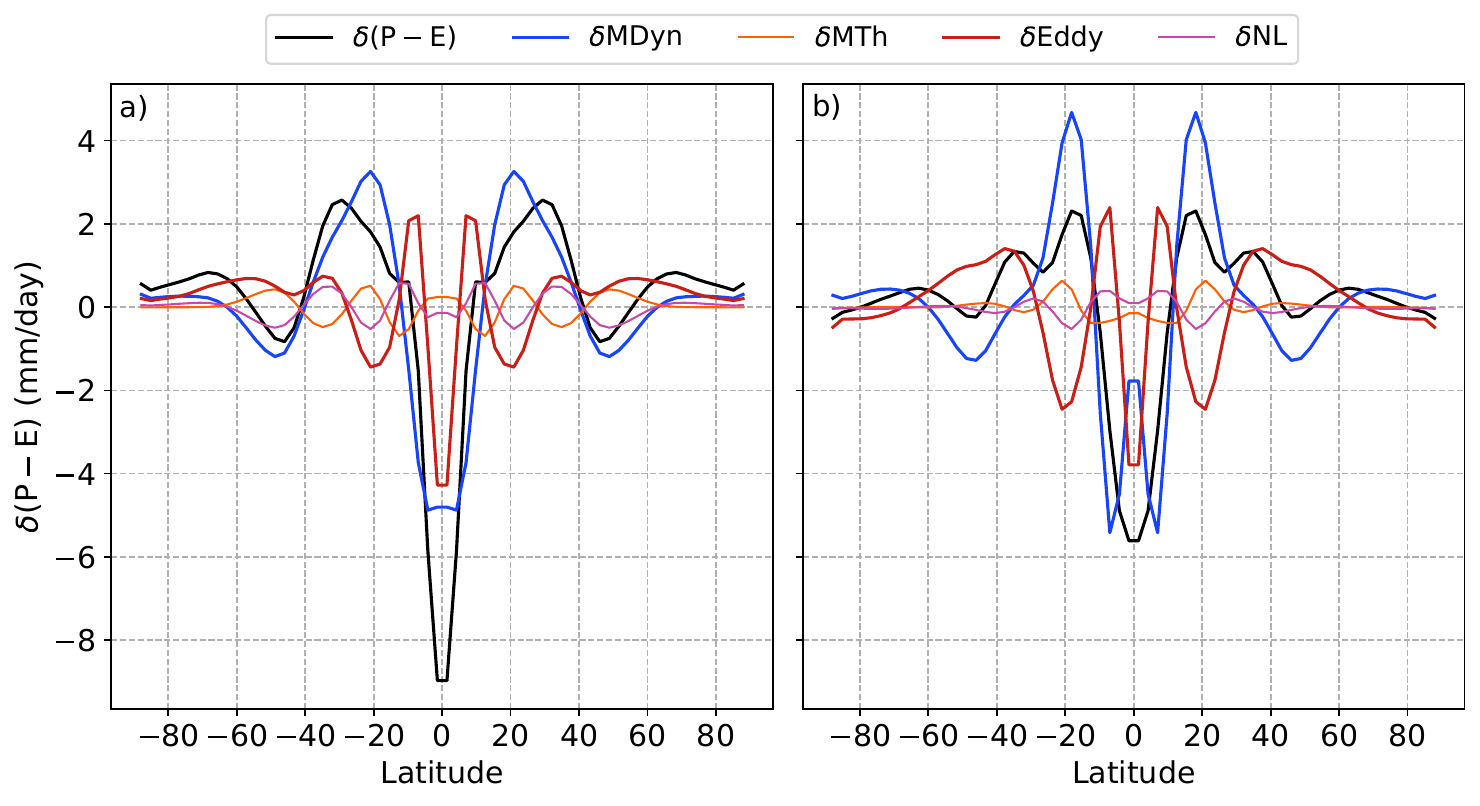}
    \caption{Decomposition of $\mathrm{P-E}$ changes [see Eq. (\ref{decomposition})] between the $F_0 = 0$ and $F_0 = 4$~m.s\textsuperscript{$-1$}.day\textsuperscript{$-1$} experiments in the interactive- (panel \textbf{a}) and fixed-SST runs (panel \textbf{b}).}
    \label{fig:P-E_diff}
\end{figure}

As shown in Figure~\ref{fig:P-E_diff}, the different components of Eq.~(\ref{decomposition}) behave differently when increasing the imposed torque.
First of all, we note that both in the fixed- and interactive-SST experiments, two components of the decomposition are clearly more important than the others: the mean dynamic component and the eddy component.
The two other components (mean thermodynamic and non-linear) are negligible.
Note that this is qualitatively different from simulations of anthropogenic climate change where the thermodynamic effect tends to dominate, leading to an approximate Clausius-Clapeyron scaling for $\mathrm{P-E}$~\citep{heldRobustResponsesHydrological2006}, which holds relatively well over ocean regions~\citep{byrneResponsePrecipitationMinus2015}.

We start by discussing the interpretation of the two dominant terms in the interactive-SST case (Fig.~\ref{fig:P-E_diff}, \textbf{a}).
The spatial structure of the total $\mathrm{P-E}$ change mostly matches the structure of the contribution due to MMC changes $\delta \mathrm{MDyn}$.
These changes are directly related to the response of the MMC to the imposed tropical torque (Fig.~\ref{fig:jet}) and are consistent with the meridional energy transport changes (Section~\ref{sec:forcing/feedback}\ref{sec:energy_transport_changes}): the collapse of the Hadley cells accounts for the negative sign of $\delta \mathrm{MDyn}$ within approximately $10^\circ$S-$10^\circ$N and its positive sign poleward of $12^\circ$ in each hemisphere.
Similarly, the large shift of the Ferrell cells contributes to the poleward flank of the region of positive moisture convergence changes between $12^\circ$-$40^{\circ}$, and explains the negative changes around $50^\circ$ as well as the small positive contribution closer to the poles.

The eddy moisture flux changes $\delta \mathrm{Eddy}$ act mostly as a modulator of the changes imposed by the MMC changes: the two components have opposite sign in the regions centered on $50^\circ$, $20^\circ$ and $10^\circ$.
This results in small shifts of the total $\mathrm{P-E}$ changes compared to the $\delta \mathrm{MDyn}$ contribution.
On the other hand, $\delta \mathrm{Eddy}$ and $\delta \mathrm{MDyn}$ have the same sign in high-latitude regions, around $30^\circ$, and at the equator; there both effects provide roughly similar contributions to the total $\mathrm{P-E}$ changes.
A natural interpretation is that the positive contribution of eddy transport to moisture convergence changes at high-latitudes can be attributed to the poleward shift of the mid-latitude jets, while the changes in the tropics result from horizontal mixing associated with eddies generated by the equatorial jet.
Indeed, the meridional structure is very similar to the structure of the changes in eddy energy transport (Fig.~\ref{fig:MSEtransport}), dominated by latent heat in this region.

The fixed-SST case (Fig.~\ref{fig:P-E_diff}, \textbf{b}) is similar to the interactive-SST case, with a number of interesting differences.
Most importantly, the spatial structure of the total $\mathrm{P-E}$ change follows less closely the MMC change contribution.
The main differences are found in the subtropics and regions poleward.
First, there remains here a much more vigorous and vertically extended meridional circulation in the subtropics, which converges moisture around 20$^\circ$ and accounts for the larger peak in $\delta \mathrm{MDyn}$ compared to the interactive-SST case. In addition, the eddy moisture transport changes are also larger than in the interactive-SST case in the 20$^\circ$-40$^\circ$ region, like for the eddy energy transport changes (Fig.~\ref{fig:MSEtransport}).

We conclude that the moisture transport changes, and therefore the precipitation changes, are a direct consequence of the circulation changes and are in particular dominated by the changes in the MMC, with a locally important contribution from eddy transport changes (at the equator and at high latitudes).

\section{Conclusion}\label{sec:conc}

In this work, we have investigated the climate changes induced by a major reorganization of the large-scale circulation in the tropical atmosphere, leading to the emergence of an eastward jet at the equator.
A transition to a state of this type, called equatorial superrotation, could occur under various circumstances leading to enhanced diabatic heating in the tropics.
Because it remains unclear under which conditions state-of-the-art coupled models can simulate such a state, and such conditions might depend on parameterizations which are not very well constrained, we have decided to circumvent the two-way coupling in a first step and ask how a transition to superrotation, forced dynamically, affects climate more generally, and in particular surface conditions.
Our main result is that even in the absence of any direct radiative forcing, the consequences of the circulation changes are major: the surface warms approximately as much as under a doubling of the atmospheric CO\textsubscript{2} concentration.
The warming is not uniform: it is particularly strong in mid-latitudes, while the tropics warm slightly less and the subtropics almost not at all.
In addition, the spatial distribution of precipitation changes dramatically, becoming more uniform over the globe, with a breakdown of the maximum associated with the ITCZ and a strong increase in the subtropics.

To unravel the mechanisms connecting the circulation changes to these temperature and precipitation changes, we have utilized a forcing/feedback analysis using radiative kernels along with decompositions of the energy and moisture transport changes into their various dynamical and thermodynamical components.
These analyses reveal that different regions of the planet respond very differently to the imposed torque.
First, the response in the tropics and subtropics is mostly governed by the moisture changes induced by the collapse of the MMC.
This results in large relative humidity changes (decreasing in the tropics and increasing in the subtropics), which allow the atmosphere to compensate radiatively the changes in meridional energy transport related to circulation changes (larger energy convergence in the tropics, smaller in the subtropics).
In other words, the radiative budget changes are dominated by the radiative forcing due to relative humidity, and feedbacks are small.
Hence, in these regions surface temperature changes are also small.
However, precipitation changes are large, and they could affect significantly the regional water cycle and therefore the biosphere.
On the other hand, in the mid- to high-latitudes, the effects of moisture transport changes are too weak to compensate the energy transport changes, and the whole atmospheric column as well as the surface must warm to equilibrate the increased horizontal energy flux convergence.
In other words, the temperature response is dominated by the Planck feedback. On a more technical note, this work demonstrates how insight can be gained into climate problems where the forcing is dynamic in origin, using various established conceptual tools (radiative forcing/feedback framework, fixed relative humidity ansatz, dry and moist EBMs, dynamic-thermodynamic decomposition of moisture transports).

Future work could investigate the robustness of the results presented here, for instance using different climate models and less idealized setups.
In particular, the role of clouds or limited moisture availability at land surfaces are likely to influence the climate responses to forced circulation changes.
In the case of clouds, previous results from~\citet{Caballero2013} suggest that they could amplify the effects described here.
It also remains to be better understood under which circumstances the climate of the Earth could spontaneously undergo such a transition, if at all, and how the two-way thermodynamic coupling between the processes driving the transition to superrotation and the impacts of superrotation described here would unfold.
Finally, if the climate changes described here prove to be robust, a natural question would be whether they can be connected to paleo proxies for warm climates of the past.

\clearpage
\acknowledgments

The authors thank the three anonymous referees for useful comments which substantially improved the manuscript.
The numerical simulations were carried out using the resources of the \emph{Pole Scientifique de Modélisation Numérique} \& \emph{Centre Blaise Pascal} at ENS de Lyon, and in particular the SIDUS system~\citep{Quemener2013}. We are grateful to Cerasela Calugaru and Emmanuel Quemener for their help with the platform.
This work was supported by the ANR Grant \emph{TippingWinds}, Project No. ANR-21-CE30-0016-01 and MPB acknowledges funding from the UK Natural Environment Research Council (Grant NE/T006269/1).
CH is also supported by the \textit{Fondation Simone et Cino Del Duca - Institut de France}. 

%
%
\datastatement

 The numerical model simulations upon which this study is based are too large to archive or to transfer.  Instead, we provide all the information needed to replicate the simulations; the model version is available online at \url{https://github.com/cbherbert/Isca/tree/eddyforcing}.  The model compilation scripts, boundary condition files and data postprocessing codes are available at \url{https://doi.org/10.5281/zenodo.15181902}.

\appendix[A] \label{appendix:radiative_kernel_q}

\appendixtitle{Radiative budget changes and their relation to temperature and specific humidity}

\begin{figure}
    \centering
    \noindent \includegraphics[width=36pc,angle=0]{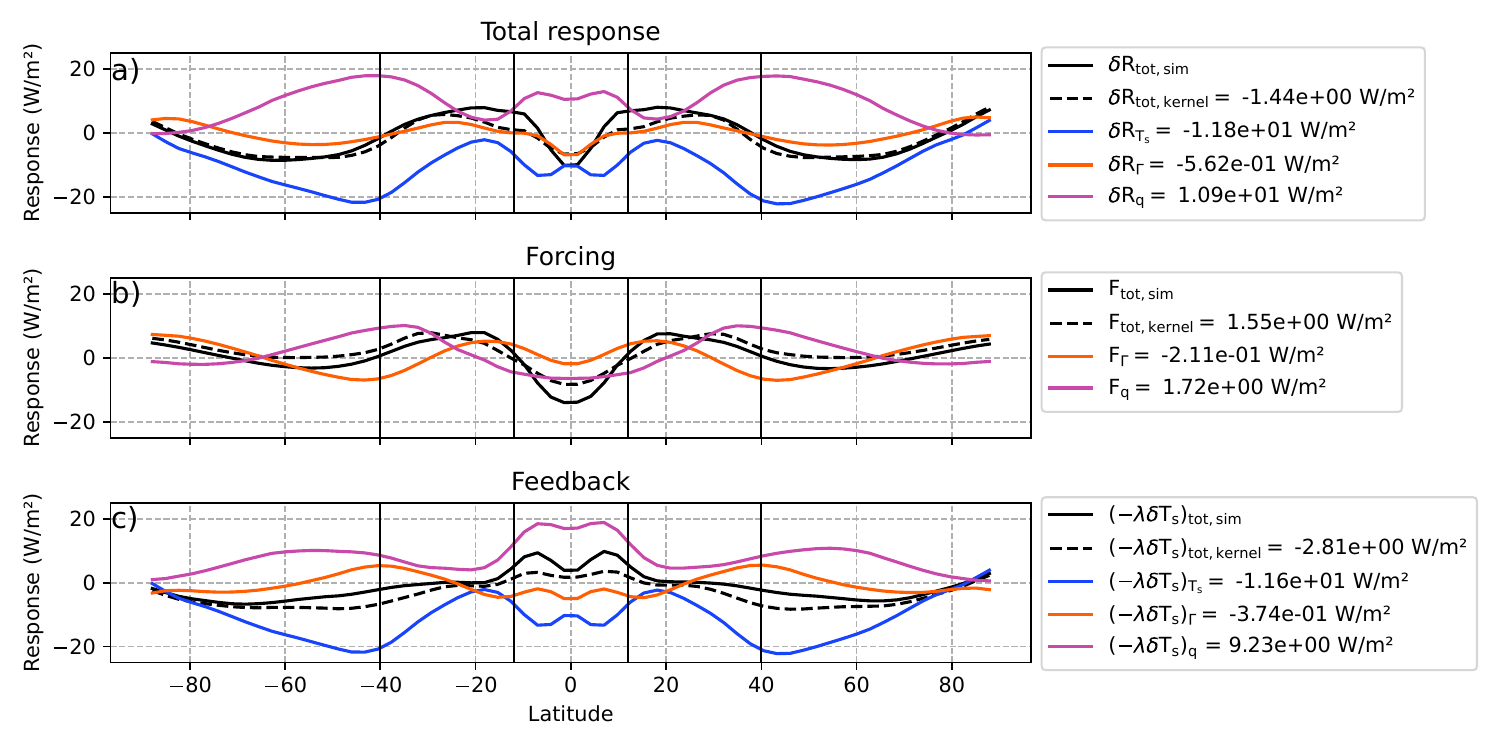}
    \caption{TOA flux changes (panel \textbf{a}) across the transition to superrotation (from $F_0 = 0$ to $F_0 = 4$~m.s\textsuperscript{$-1$}.day\textsuperscript{$-1$}), and the decomposition of these changes into radiative forcings (panel \textbf{b}) and feedbacks (panel \textbf{c}) computed with radiative kernel.
      Panel \textbf{a} corresponds to the interactive-SST runs, panel \textbf{b} to the fixed-SST simulations and panel \textbf{c} is the difference between the two.
      Black curves represent the radiative budgets computed by the radiation scheme in the model, blue curves are the budgets reconstructed using the radiative kernels, which can be broken down into the Planck response (yellow curves), specific humidity response (red curves) and lapse-rate response (green curves).
    Numerical values given in the legend correspond to the global averages of the various components.}
    \label{fig:decomp_q}
\end{figure}

As an alternative to the relative humidity-based decomposition used in Section~\ref{sec:forcing/feedback}\ref{sec:radiative_kernel_rh}, we can analyze the TOA radiative budget changes using the traditional decomposition~\citep{sodenQuantifyingClimateFeedbacks2008} where specific humidity is an independent variable (in addition to lapse rate and surface temperature).
As shown in Fig.~\ref{fig:decomp_q}, in this alternative decomposition, the results are more difficult to interpret because all terms have non-negligible contributions in all regions.
Nevertheless, relating changes in TOA fluxes to changes in temperature and specific humidity brings additional insight into the mechanisms at play.

\begin{figure}
  \centering
  \noindent \includegraphics[width=36pc]{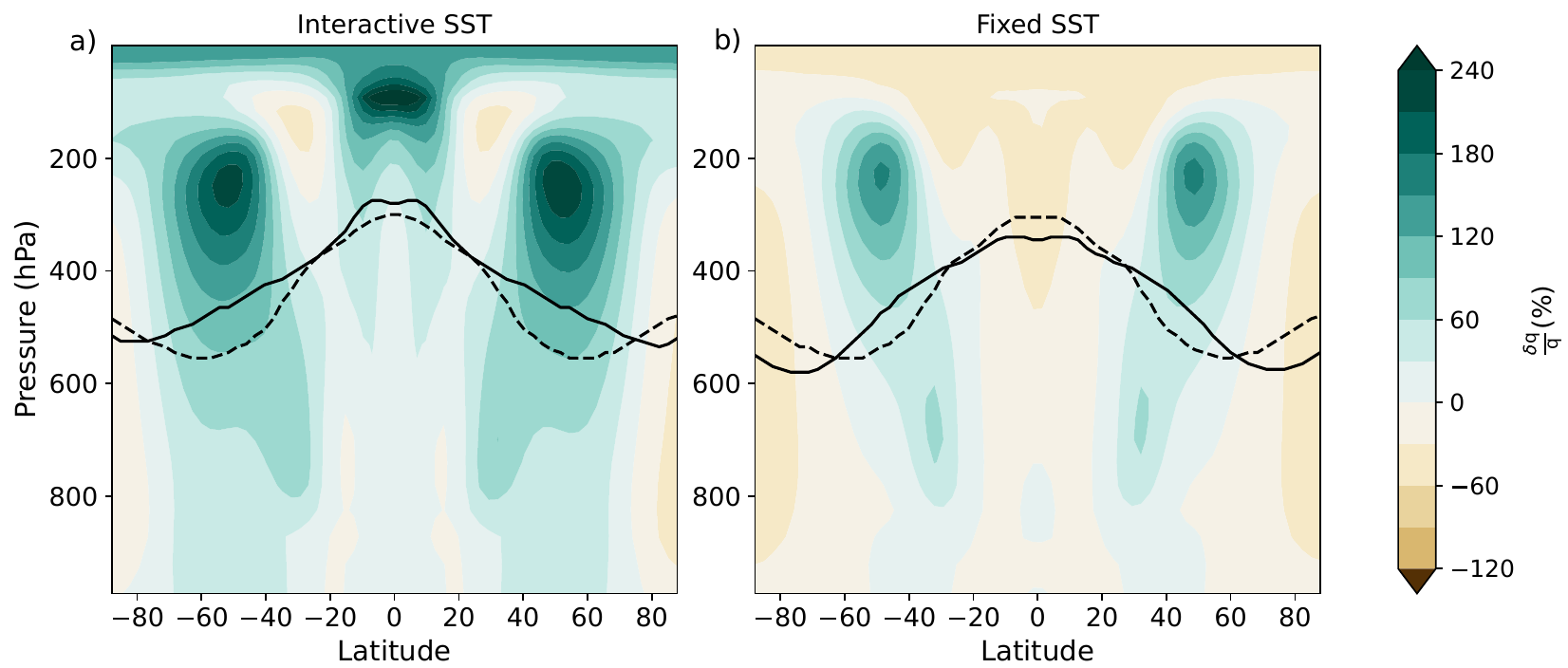}
  \caption{Time- and zonally-averaged fractional specific humidity changes between the $F_0~=~4$~m.s\textsuperscript{$-1$}.day\textsuperscript{$-1$} and $F_0~=~0$~m.s\textsuperscript{$-1$}.day\textsuperscript{$-1$} experiments with interactive (panel \textbf{a}) and fixed SST (panel \textbf{b}).
    Here and for Figure~\ref{fig:T_emission_height}, the dashed and solid black lines show the time- and zonally-averaged emission heights for $F_0 = 0$ and $F_0 = 4$~m.s\textsuperscript{$-1$}.day\textsuperscript{$-1$}, respectively. }
  \label{fig:q_emission_height}
\end{figure}

\begin{figure}
  \centering
  \noindent \includegraphics[width=36pc]{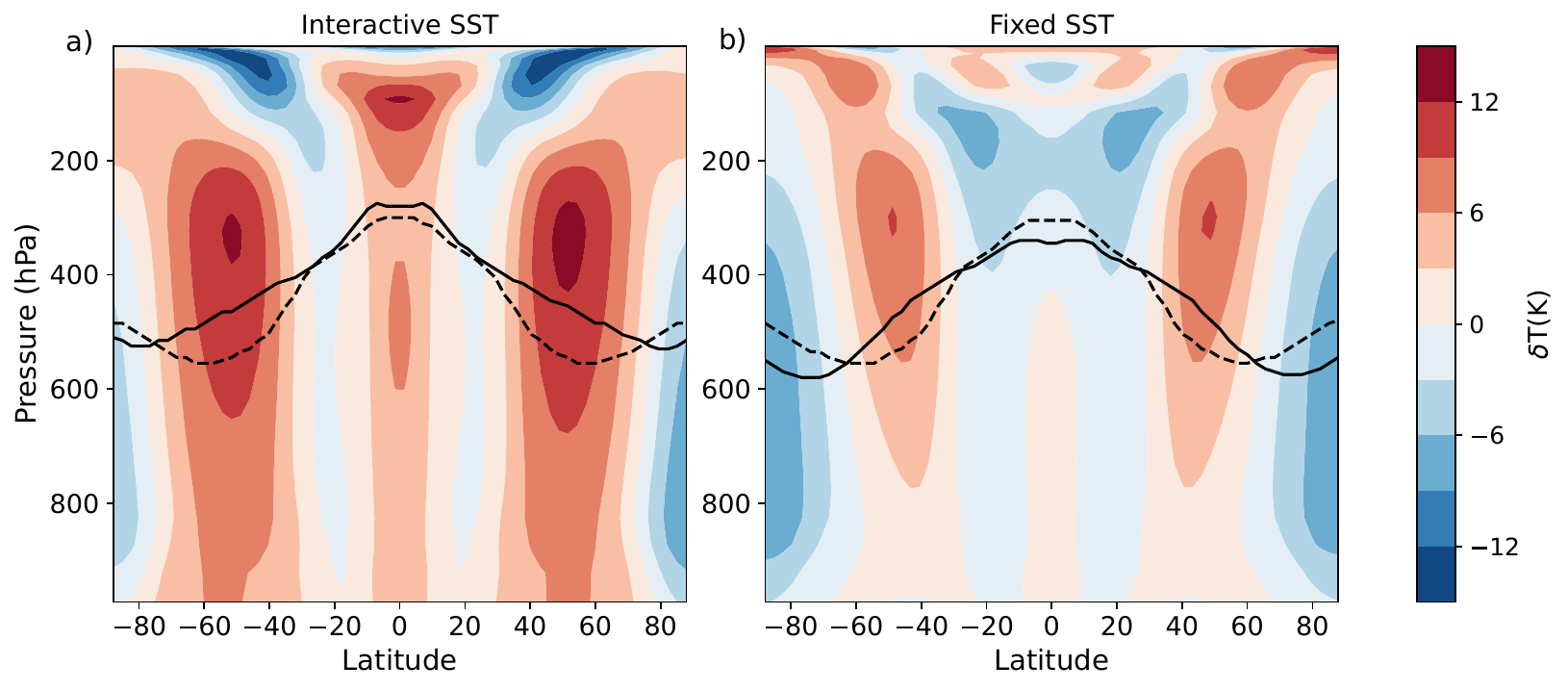}
  \caption{Time- and zonally-averaged atmospheric temperature changes between the $F_0~=~4$~m.s\textsuperscript{$-1$}.day\textsuperscript{$-1$} and $F_0~=~0$~m.s\textsuperscript{$-1$}.day\textsuperscript{$-1$} experiments with interactive (panel \textbf{a}) and fixed SST (panel \textbf{b}).}
  \label{fig:T_emission_height}
\end{figure}

We start with the meridional structure of the radiative forcing (i.e., the TOA anomaly induced in the fixed-SST runs by prescribing $F_0 = 4$~m.s\textsuperscript{$-1$}.day\textsuperscript{$-1$}).
We show in Figure~\ref{fig:q_emission_height} (panel \textbf{b}) and Figure~\ref{fig:T_emission_height} (panel \textbf{b}) the time- and zonally-averaged specific humidity and temperature changes, respectively.
In the superrotating state, the atmosphere is drier than the control in the tropics, moister in the mid-latitudes, and slightly drier close to the poles.
This explains qualitatively the meridional structure of the contribution of specific humidity changes to the radiative forcing (Fig.~\ref{fig:decomp_q}, panel \textbf{b}): drier air results in decreased optical depth, increased OLR, and negative radiative forcing, all else equal (conversely for moister air).
Similarly, Figure~\ref{fig:T_emission_height} (panel \textbf{b}) shows that the lapse rate increases in the tropics, decreases in mid-latitudes, and increases again close to the poles, which matches the meridional structure of the lapse rate contribution to radiative forcing (Fig.~\ref{fig:decomp_q}, panel \textbf{b}).

However, the way the two effects combine into the net radiative forcing is more subtle (Fig.~\ref{fig:decomp_q}, panel \textbf{b}).
To understand it, we consider jointly changes in the average emission level due to specific humidity changes and the spatial distribution of temperature changes (Figs.~\ref{fig:q_emission_height} and ~\ref{fig:T_emission_height}). Note that the emission level is defined as $p_{\rm em}(\varphi) = \left\{p(\varphi) \text{ such that } T(p,\varphi) = \left( \frac{\rm \overline{OLR}(\varphi)}{\sigma} \right)^{1/4} \right\} $, where $\sigma$ is the Stefan-Boltzmann constant.
Close to the equator (below $12^{\circ}$ latitude), the drier atmosphere results in a substantially lower emission level, which is not compensated for by the lapse rate changes.
Although the temperature in the superrotating state at the new emission level is cooler than it was at the same level in the control run, it is warmer than the temperature at the emission level in the control run, hence the OLR is larger: specific humidity effects dominate.
Moving poleward, the competition between the two effects progressively changes: while the atmosphere is still drier in the superrotating state, it is less so and the emission level changes are smaller.
At the same time, the lapse rate changes amplify.
The combined change results in decreased OLR changes, until it vanishes around 12$^\circ$ latitude and even becomes negative when lapse rate effects dominate between $12^\circ$ and $20^\circ$.
Between $20^\circ$ and $40^{\circ}$, the two effects reverse (Fig.~\ref{fig:T_emission_height}): the lapse rate changes decrease until they become negative around $30^\circ$ latitude and poleward, and the atmosphere progressively becomes moister than the control run (Fig.~\ref{fig:q_emission_height}), resulting in higher emission levels.
As a result, the lapse rate contribution to the radiative forcing decreases with latitude in this region (Fig.~\ref{fig:decomp_q}, panel \textbf{b}) while the specific humidity contribution increases before levelling off.
The fact that the OLR is smaller than in the control run in this region (the blue curve is positive) is therefore driven first by the specific humidity effects in the first half of the region, then by lapse rate effects in the second half.
Going further poleward, the two effects essentially cancel out, resulting in a vanishing radiative forcing up to very high latitudes.

For the interactive-SST experiments, Figure~\ref{fig:q_emission_height} (panel \textbf{a}) shows that the specific humidity changes are smaller than in the fixed-SST case equatorward of $20^\circ$ (in this region, the emission level remains the same) --- the superrotating state is actually moister except in the region where the OLR value is larger.
Lapse rate changes are essentially smaller everywhere (Fig.~\ref{fig:T_emission_height}, panel \textbf{a}).
This explains the structure of the radiative feedback curves (Fig.~\ref{fig:decomp_q}, panel \textbf{c}), whose sign is opposite to the radiative forcing everywhere for the lapse rate component (green curve), and opposite in the tropics but the same sign in mid-latitudes for the specific humidity component (red curve; in other words specific humidity acts as a positive feedback everywhere).
As a result, the specific humidity effect on changes in the net radiative budget vanishes in the tropics while the lapse rate component remains relatively small everywhere (Fig.~\ref{fig:decomp_q}, panel \textbf{a}).
The total radiative budget changes are dominated by the Planck response in the deep tropics (with a minor contribution from the lapse rate), and become dominated by the imperfect cancellation between the Planck and specific humidity responses at higher latitudes.

Overall, the TOA radiative budget changes are consistent with changes in physical quantities in the atmosphere, such as specific humidity and temperature, but result from cancellations between the effects associated with these variables.
As shown in section~\ref{sec:forcing/feedback}\ref{sec:radiative_kernel_rh}, these cancellations indicate that the effects of specific humidity and lapse rate on the radiative budget are essentially a consequence of the Clausius-Clapeyron relation.
Hence, the radiative budget changes are better described by changes in relative humidity, which we use in the main text as a state variable instead of specific humidity.

\appendix[B] \label{appendix:rk_derivation}

\appendixtitle{Derivation of the relative humidity radiative response}

We derive the expression for the relative humidity-based decomposition of TOA radiative budget changes (introduced in Section~\ref{sec:forcing/feedback}\ref{sec:radiative_kernel_rh}), starting from the specific humidity-based decomposition:
\begin{align}
    \nonumber \delta R & = \delta R (T_{\mathrm{s}}, T, q) \approx  \left.\frac{\partial R}{\partial T_{\mathrm{s}}}\right|_{T,q} \delta T_{\mathrm{s}} + \left.\frac{\partial R}{\partial T}\right|_{T_{\mathrm{s}},q} \delta T + \left.\frac{\partial R}{\partial q}\right|_{T_{\mathrm{s}},T} \delta q \\
    & \approx \left.\frac{\partial R}{\partial T_{\mathrm{s}}}\right|_{T,q} \delta T_{\mathrm{s}} + \left.\frac{\partial R}{\partial T}\right|_{T_{\mathrm{s}},q} \delta T + \left.\frac{\partial R}{\partial q}\right|_{T_{\mathrm{s}},T} \left( q^{\star} \delta \mathrm{RH} + \mathrm{RH}  \frac{\partial q^{\star}}{\partial T} \delta T \right) \\
    \nonumber & \approx \left( \left.\frac{\partial R}{\partial T}\right|_{T_{\mathrm{s}},q} + \left.\frac{\partial R}{\partial q}\right|_{T_{\mathrm{s}},T} \frac{\partial q}{\partial T}\right) \delta T + \left.\frac{\partial R}{\partial T_{\mathrm{s}}}\right|_{T,q}  \delta T_{\mathrm{s}} + \left.\frac{\partial R}{\partial q}\right|_{T_{\mathrm{s}},T} q^{\star} \delta \mathrm{RH}.
\end{align}

To split the atmospheric temperature change into a vertically-uniform component (the Planck response) and a vertically-resolved anomaly (the lapse rate response), we can write $\delta T = \delta T_{\mathrm{s}} + \delta (T-T_{\mathrm{s}})$ and we therefore have:
\begin{align}
    \nonumber \delta R \approx & \left( \left.\frac{\partial R}{\partial T}\right|_{T_{\mathrm{s}},q} + \left.\frac{\partial R}{\partial q}\right|_{T_{\mathrm{s}},T} \frac{\partial q}{\partial T}\right) \delta (T-T_{\mathrm{s}}) \\
    & + \left(  \left.\frac{\partial R}{\partial T}\right|_{T_{\mathrm{s}},q} + \left.\frac{\partial R}{\partial q}\right|_{T_{\mathrm{s}},T} \frac{\partial q}{\partial T} + \left.\frac{\partial R}{\partial T_{\mathrm{s}}}\right|_{T,q}\right) \delta T_{\mathrm{s}}\\
    &\nonumber  + \left.\frac{\partial R}{\partial q}\right|_{T_{\mathrm{s}},T} q^{\star} \delta \mathrm{RH}.
\end{align}

The radiative kernels are expressed as:
\begin{equation}
    \left\{ \begin{array}{c}
         K^{T_{\mathrm{s}}}  = \left.\frac{\partial R}{\partial T_{\mathrm{s}}}\right|_{T,q}  \\
         K^{T} = \left.\frac{\partial R}{\partial T}\right|_{T_{\mathrm{s}},q}\\
         K^{\mathrm{wv}} = \left.\frac{\partial R}{\partial \log(q)}\right|_{T_{\mathrm{s}},T} \frac{\partial \log(q)}{\partial T}.
    \end{array} \right.
\end{equation}

Since $\left. \frac{\partial R}{\partial q} \right|_{T_{\mathrm{s}},T} q^{\star} = \left.\frac{\partial R}{\partial \log(q)}\right|_{T_{\mathrm{s}},T} \frac{1}{\mathrm{RH}} $, we can change the state variables and write:
\begin{align}
    \begin{pmatrix}
        \delta R_q \\
        \delta R_{T_\mathrm{s}} \\
        \delta R_{\Gamma}
    \end{pmatrix} = \begin{pmatrix}
        K^{\mathrm{wv}} \frac{\delta \log(q)}{\partial \log (q) / \partial T} \\
        \left( K^{T_{\mathrm{s}}} + K^{T} \right) \delta T_{\mathrm{s}} \\
        K^T \left( \delta T - \delta T_{\mathrm{s}} \right)
    \end{pmatrix} \xrightarrow[\mathrm{{state \, variables}}]{\mathrm{Change \, of}} \begin{pmatrix}
        \widetilde{\delta R}_{\mathrm{RH}} \\
        \widetilde{\delta R}_{T_{\mathrm{s}}} \\
        \widetilde{\delta R}_{\Gamma}
    \end{pmatrix} = \begin{pmatrix}
        K^{\mathrm{wv}} \frac{\delta \log\left(\mathrm{RH}\right)}{\partial \log(q) / \partial T} \\
        \left( K^{T_{\mathrm{s}}} + K^{T} + K^{\mathrm{wv}} \right) \delta T_{\mathrm{s}} \\
        \left( K^T + K^{\mathrm{wv}} \right) \left( \delta T - \delta T_{\mathrm{s}} \right)
    \end{pmatrix},
\end{align}
where the tildes indicate the different choice of independent variables (the tildes are not included in the main text).

We note that the relative-humidity component of $\widetilde{\delta R}$ can be written as:
\begin{align}
    \nonumber \widetilde{\delta R}_{\mathrm{RH}} = \left.\frac{\partial R}{\partial \log(q)}\right|_{T_{\mathrm{s}},T} \delta\log(\mathrm{RH}) &= \left.\frac{\partial R}{\partial \log(q)}\right|_{T_{\mathrm{s}},T} \frac{1}{\mathrm{RH}} \frac{1}{q^ {\star}} \left( \delta q - \mathrm{RH} \delta q^{\star} \right)\\
    & = K^{\mathrm{wv}} \frac{\delta\log(q)}{\partial\log(q)/\partial T} - \left.\frac{\partial R}{\partial \log(q)}\right|_{T_{\mathrm{s}},T} \frac{1}{q^{\star}} \frac{\partial q^{\star}}{\partial T} \delta T \\
    \nonumber & =K^{\mathrm{wv}} \frac{\delta\log(q)}{\partial\log(q)/\partial T} - K^{\mathrm{wv}} \delta T,
\end{align}
which closes the description of the method used for the computations presented
in Section~\ref{sec:forcing/feedback}\ref{sec:radiative_kernel_rh}.


\bibliographystyle{ametsocV6}
\bibliography{biblio}

\end{document}